\documentclass{aa}  
\def \eg {e.g.}

\def \ie {i.e.}
\def \cf {cf.}
\def \lcdm {{\hbox{$\Lambda$CDM}}}
\def \omegam {{\hbox{$\Omega_m$}}}
\def \omegal {{\hbox{$\Omega_\Lambda$}}}
\def \hzero {{\hbox{$H_0$}}}
\def \arcmin {\hbox{$^\prime$}}
\def \arcsec {\hbox{$^{\prime\prime}$}}
\def \deg {\hbox{$^\circ$}}
\def \nh {\hbox{$N_{\rm H}$}}
\def \ktu {{\hbox{$kT_u$}}}
\def \ktd {{\hbox{$kT_d$}}}
\def \mach {{\hbox{$\mathcal{M}$}}}
\def \machkt {{\hbox{$\mathcal{M}_{\rm kT}$}}}
\def \machsb {{\hbox{$\mathcal{M}_{\rm SB}$}}}
\def \compr {{\hbox{$\mathcal{C}$}}}

\def \msun {\hbox{${\rm M_\odot}$}}

\def \mfive {\hbox{$M_{500}$}}

\newcommand{\kmsmpc }{\mbox{km s$^{-1}$ Mpc$^{-1}$}}

\newcommand{\kev }{\mbox{keV}}

\newcommand{\jy }{\mbox{Jy}}

\newcommand{\mujyb }{\mbox{$\mu$Jy beam$^{-1}$}}

\newcommand{\whz }{\mbox{W Hz$^{-1}$}}

\newcommand{\epic }{EPIC}
\newcommand{\epicE }{European Photon Imaging Camera}
\newcommand{\obsid }{ObsID}

\newcommand{\uv }{\textit{uv}}

\newcommand{\factor }{\textsc{factor}}
\newcommand{\prefactor }{\textsc{prefactor}}

\newcommand{\wsclean }{\textsc{WSClean}}

\newcommand{\spam }{\textsc{spam}}
\newcommand{\spamE }{Source Peeling and Atmospheric Modeling}
\newcommand{\pybdsf }{\textsc{pybdsf}}
\newcommand{\pybdsfE }{PYthon Blob Detector and Source Finder}

\newcommand{\xspec }{\textsc{xspec}}

\newcommand{\esas }{\textsc{esas}}
\newcommand{\esasE }{Extended Source Analysis Software}
\newcommand{\sas }{\textsc{sas}}
\newcommand{\sasE }{Scientific Analysis System}
\newcommand{\proffit }{\textsc{proffit}}
\newcommand{\xmm }{{\em XMM-Newton}}

\newcommand{\spitzer }{{\em Spitzer}}

\newcommand{\planck }{{\em Planck}}

\newcommand{\gmrt }{GMRT}
\newcommand{\gmrtE }{Giant Metrewave Radio Telescope}

\newcommand{\vla }{VLA}
\newcommand{\vlaE }{Very Large Array}

\newcommand{\lofar }{LOFAR}
\newcommand{\lofarE }{LOw Frequency ARray}
\newcommand{\ska }{SKA}
\newcommand{\skaE }{Square Kilometre Array}
\newcommand{\mwa }{MWA}
\newcommand{\mwaE }{Murchison Widefield Array}

\newcommand{\lotss }{LoTSS}
\newcommand{\lotssE }{LOFAR Two-metre Sky Survey}

\newcommand{\first }{FIRST}
\newcommand{\firstE }{Faint Images of the Radio Sky at Twenty-cm}
\newcommand{\nvss }{NVSS}
\newcommand{\nvssE }{NRAO VLA Sky Survey}
\newcommand{\wenss }{WENSS}
\newcommand{\wenssE }{WEsterbork Northern Sky Survey}

\newcommand{\sdss }{SDSS}
\newcommand{\sdssE }{Sloan Digital Sky Survey}

\newcommand{\xcop }{X-COP}
\newcommand{\xcopE }{{\em XMM-Newton} Cluster Outskirts Project}
\usepackage{graphicx}
\usepackage{amssymb}
\usepackage{multirow,bigdelim}
\usepackage{txfonts}
\usepackage[export]{adjustbox}
\defcitealias{govoni11}{G11}
\defcitealias{venturi11}{V11}

\begin{document} 

\title{The spectacular cluster chain Abell 781 as observed with \lofar, \gmrt, and \xmm}

\authorrunning{A. Botteon et al.} 
\titlerunning{\lofar, \gmrt\ and \xmm\ observations of the cluster chain Abell 781}

\author{A. Botteon\inst{1,2}, T. W. Shimwell\inst{3,4}, A. Bonafede\inst{1,2,5}, D. Dallacasa\inst{1,2}, F. Gastaldello\inst{6}, D. Eckert\inst{7}, G. Brunetti\inst{2}, T. Venturi\inst{2}, R. J. van Weeren\inst{4}, S. Mandal\inst{4}, M. Br\"{u}ggen\inst{5}, R. Cassano\inst{2}, F. de Gasperin\inst{4,5}, A. Drabent\inst{8}, C. Dumba\inst{8,9}, H. T. Intema\inst{4}, D. N. Hoang\inst{4}, D. Rafferty\inst{5}, H. J. A. R\"{o}ttgering\inst{4}, F. Savini\inst{5}, A. Shulevski\inst{10}, A. Stroe\inst{11} \and A. Wilber\inst{5}}

\institute{
Dipartimento di Fisica e Astronomia, Universit\`{a} di Bologna, via P.~Gobetti 93/2, I-40129 Bologna, Italy \\
\email{botteon@ira.inaf.it}
\and
INAF - IRA, via P.~Gobetti 101, I-40129 Bologna, Italy
\and
ASTRON, the Netherlands Institute for Radio Astronomy, Postbus 2, NL-7990 AA Dwingeloo, The Netherlands
\and
Leiden Observatory, Leiden University, PO Box 9513, NL-2300 RA Leiden, The Netherlands
\and
Hamburger Sternwarte, Universit\"{a}t Hamburg, Gojenbergsweg 112, D-21029 Hamburg, Germany
\and
INAF - IASF Milano, via E.~Bassini 15, I-20133 Milano, Italy
\and
Max Planck Institut f\"{u}r Extraterrestrische Physik, Giessenbachstrasse 1, D-85748 Garching, Germany
\and
Th\"{u}ringer Landessternwarte, Sternwarte 5, D-07778 Tautenburg, Germany
\and
Mbarara University of Science \& Technology, PO Box 1410 Mbarara, Uganda 
\and
Anton Pannekoek Institute for Astronomy, University of Amsterdam, Postbus 94249, NL-1090 GE Amsterdam, The Netherlands
\and
European Southern Observatory, Karl-Schwarzschild-Str. 2, D-85748 Garching, Germany 
}

\date{Received XXX; accepted YYY}

\abstract
{A number of merging galaxy clusters show the presence of large-scale radio emission associated with the intra-cluster medium (ICM). These synchrotron sources are generally classified as radio haloes and radio relics.}
{Whilst it is commonly accepted that mergers play a crucial role in the formation of radio haloes and relics, not all the merging clusters show the presence of giant diffuse radio sources and this provides important information concerning current models. The Abell 781 complex is a spectacular system composed of an apparent chain of clusters on the sky. Its main component is undergoing a merger and hosts peripheral emission that is classified as a candidate radio relic and a disputed radio halo.}
{We used new LOw Frequency ARay (LOFAR) observations at 143 MHz and archival Giant Metrewave Radio Telescope (GMRT) observations at 325 and 610 MHz to study radio emission from non-thermal components in the ICM of Abell 781. Complementary information came from \xmm\ data, which allowed us to investigate the connection with the thermal emission and its complex morphology.}
{The origin of the peripheral emission is still uncertain. We speculate that it is related to the interaction between a head tail radio galaxy and shock. However, the current data allow us only to set an upper limit of $\mach<1.4$ on the Mach number of this putative shock. Instead, we successfully characterise the surface brightness and temperature jumps of a shock and two cold fronts in the main cluster component of Abell 781. Their positions suggest that the merger is involving three substructures. We do not find any evidence for a radio halo either at the centre of this system or in the other clusters of the chain. We place an upper limit to the diffuse radio emission in the main cluster of Abell 781 that is a factor of 2 below the current radio power-mass relation for giant radio haloes.}
{}


\maketitle
%

\section{Introduction}

\begin{figure*}
 \centering
 \includegraphics[width=.75\textwidth,trim={0cm 0cm 0cm 0.2cm},clip]{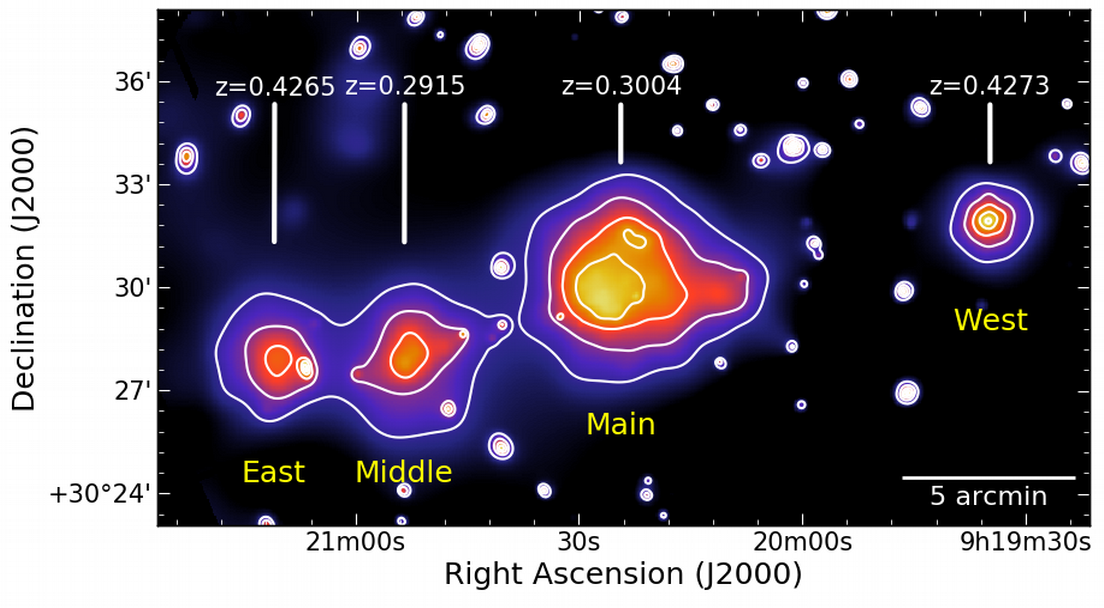}
 \caption{Adaptively smoothed, background-subtracted, and exposure-corrected \xmm\ mosaic image in the $0.5-2.0$ \kev\ band of the Abell 781 complex. Contours are spaced by a factor of 2 starting from $3.5\times10^{-6}$ counts s$^{-1}$ pixel$^{-1}$.}
 \label{fig:smoothed}
\end{figure*}

Mergers between galaxy clusters are the most energetic phenomena in the Universe after the Big Bang with the total kinetic energy of the collision reaching $10^{64}$ erg in a crossing timescale ($\sim$ Gyr). Part of this energy is dissipated in non-thermal processes in the intra-cluster medium (ICM), \ie\ in the (re)acceleration of relativistic particles and amplification of the magnetic field \citep[\eg][for reviews]{dolag08rev, brunetti14rev}. The observation of diffuse synchrotron radio emission in some merging galaxy clusters probes the presence of non-thermal components spread on Mpc-scale in the ICM. These non-thermal sources are characterised by steep spectra (\ie\ $\alpha>1$, with $S_\nu \propto \nu^{-\alpha}$), and are commonly classified as radio haloes and radio relics \citep[\eg][for and observational overview]{feretti12rev}. \\
\indent
Radio haloes are generally centrally located, show roundish morphologies roughly tracing the X-ray thermal emission, and are apparently unpolarised. According to the favoured scenario, radio haloes are produced by the turbulence injected into the ICM during major merger events \citep{brunetti01coma, petrosian01}. This model seems to be supported by the connection observed between clusters hosting radio haloes and dynamically disturbed systems \citep[\eg][]{buote01, cassano10connection, cuciti15}. Nonetheless, the details of the mechanisms that channel the energy released on large scales to collisionless small scales in the ICM are still poorly understood \citep[\eg][]{brunetti11mfp, miniati15run, brunetti16challenge}. \\
\indent
Radio relics are polarised and elongated sources that are usually found in the outskirts of galaxy clusters. Shocks generated in cluster mergers have been proposed to explain the origin of radio relics \citep{ensslin98relics, roettiger99a3667}. Indeed, the relic--shock connection is supported by the increasing number of shocks detected at the location of radio relics \citep[\eg][for recent works]{botteon16gordo, eckert16a2744, akamatsu17a2255}. However, the luminosity of some radio relics is much higher than expected from the presumed acceleration efficiency of thermal electrons owing to diffuse shock acceleration (DSA) at low Mach number ($\mach < 3$) shocks (see \citealt{brunetti14rev} and \citealt{botteon16a115, eckert16a2744, vanweeren16toothbrush, hoang18a1240} for more recent papers). In this respect, the presence of an existing population of relativistic electrons to re-accelerate is a prerequisite of some more  recent models \citep[\eg][]{kang11, kang16reacc, pinzke13, caprioli18} that seems to be corroborated by a number of observations \citep[\eg][]{bonafede14reacc, shimwell15, vanweeren17a3411}. \\
\indent
Observations at low frequencies with the \lofarE\ \citep[\lofar;][]{vanhaarlem13}, \mwaE\ \citep[\mwa;][]{tingay13}, and, in the future, with the \skaE\ \citep[\ska;][]{dewdney09}, are expected to detect many new diffuse radio sources in galaxy clusters that can be used to increase our knowledge of non-thermal phenomena in the ICM \citep[\eg][]{rottgering06, rottgering11, cassano10lofar, nuza12}. In particular, the \lotssE\ \citep[\lotss;][]{shimwell17} is observing the northern sky at $120-168$~MHz and will produce images with unprecedented resolution ($\sim 5$ \arcsec) and sensitivity ($\sim 100$ \mujyb) in this frequency range. \\
\indent
Abell 781 is a complex system with multiple galaxy cluster components \citep{wittman06, wittman14, abate09, geller10, cook12}. In X-ray wavelengths, it appears as a chain with four prevailing clusters that extends over $\sim25\arcmin$ in the E-W direction \citep{wittman06, sehgal08}. Fig.~\ref{fig:smoothed} shows an \xmm\ image of the system where we labelled the clusters following \citet{sehgal08} and reported the redshifts from \citet{geller10}. These four clusters lie in two different redshift planes: the ``Main'' (hereafter referred to as A781) and ``Middle'' are located at $z\sim0.30,$ whereas the ``East'' and ``West'' are located at $z\sim0.43$; therefore they are not related to the other two clusters of the system. The mass of the main cluster is $\mfive = (6.1\pm0.5) \times 10^{14}$ \msun, as reported in the second \planck\ catalogue of Sunyaev-Zel'dovich sources \citep[PSZ2;][]{planck16xxvii}. \\
\indent
Observations taken with the \gmrtE\ (\gmrt) at 610~MHz revealed the presence of a peripheral source at the boundary of the X-ray thermal emission of A781, which was suggested to be a candidate radio relic by \citet{venturi08}. Although this interpretation would be in agreement with the location of the emission in the cluster outskirts, the source morphology is puzzling: neither arc-like nor elongated, its morphology changes from 610 to 325 MHz \citep[][hereafter \citetalias{venturi11}]{venturi11}. The source is also detected with the \vlaE\ (\vla) at 1.4~GHz \citep[][hereafter \citetalias{govoni11}]{govoni11}. The presence of a central radio halo in A781 is also disputed, it was observed at high frequency with the \vla\ \citepalias{govoni11} but not at lower frequencies with the \gmrt\ \citep{venturi08, venturi11, venturi13}. \\
\indent
In this work, we present a new \lofar\ observation at $120-168$~MHz and the reanalysis of archival \gmrt\ and \xmm\ observations of the cluster chain Abell 781. In particular, we focus on the main merging cluster of the complex to study the peripheral source and shed light on the presence of the radio halo that has been reported in the literature. \\
\indent
Throughout the paper, we assume a \lcdm\ cosmology with $\omegal = 0.7$, $\omegam = 0.3$ and $\hzero = 70$ \kmsmpc, for which the luminosity distance is $D_L = 1555$ Mpc and 1\arcsec\ equals $4.458$ kpc at the redshift of A781 ($z=0.3004$). Uncertainties are provided at the $1\sigma$ confidence level for one parameter, unless stated otherwise.

\section{Observations and data reduction}

\begin{table*}
 \centering
 \caption{Summary of the radio observations used in this work.}
 \label{tab:radio_obs}
  \begin{tabular}{lcccc} 
  \hline
   & \lofar\ & \gmrt\ & \gmrt\ \\
  \hline
  Project code & LC6\_015 & 11TVA01 & 08RCA01 \\
  Observation date & 2016 Dec 02 & 2007 Jan 29 & 2005 Oct 02 \\
  Total on-source time (hr) & 8.0 & 9.2 & 3.4 \\
  Flux calibrator & 3C196 & 3C286 & 3C48 \\
  Total on-calibrator time (min) & 10 & 30 & 34 \\
  Central frequency (MHz) & 143 & 325 & 610  \\
  Bandwidth (MHz) & 48 & 33 & 33 \\
  \hline
  \end{tabular}
\end{table*}

\subsection{\lofar}\label{sec:lofar}

We analysed the \lotss\ \citep{shimwell17, shimwell18} pointing closest to A781 (offset by $\sim 1.5\deg$). The observation is 8~hr long and used the Dutch High Band Antenna (HBA) array operating at $120-168$~MHz (see Tab.~\ref{tab:radio_obs} for more details). \\
\indent
The data reduction performed in this work follows the facet calibration scheme developed to analyse \lofar\ HBA data \citep{vanweeren16calibration, williams16, degasperin18systematic}. This procedure consists of two steps, and can be summarised as follows. \\

\begin{enumerate}
 \item In the first step, direction-independent calibration is performed via \prefactor\footnote{\url{https://github.com/lofar-astron/prefactor}}. Data are averaged and flagged to reduce the data set size and excise bad quality data. The flux density calibrator 3C196 is used to calibrate complex gains and clock offsets between different antenna stations adopting the absolute flux density scale of \citet{scaife12}. After the transfer of amplitude and clock solutions to the target, an initial phase calibration is performed using a Global Sky Model for \lofar\footnote{\url{https://support.astron.nl/LOFARImagingCookbook}}. This is followed by a preliminary low- and high-resolution imaging of the entire field of view (FOV). Compact and diffuse sources are detected in these images with the \pybdsfE\ (\pybdsf; \citealt{mohan15}) and then subtracted from the \uv-data creating ``blank'' field data sets in preparation of the next step of the data reduction.
 \item In the second step, direction-dependent calibration is performed via \factor\footnote{\url{https://github.com/lofar-astron/factor}}. The large primary beam of \lofar\ requires that phase and amplitude calibration solutions are computed in small portions of the sky because of the different distortions introduced by the ionosphere and because of the beam model errors over the FOV. For this reason, the FOV is tessellated in facets where a facet calibrator (generally a bright source or a group of closely spaced sources) is used to evaluate the gain and phase solutions in a restricted area of the sky. A number of self-calibration cycles are performed on the facet calibrator, then its solutions are used to calibrate the faint sources subtracted in the previous step that are added back to the data after the self-calibration of the facet. Before moving to the subsequent facet,which generally has a fainter flux density calibrator, the clean components of the processed facet are subtracted from the \uv- data to reduce the systematics and effective noise in the data set. This procedure is repeated for all the directions leaving the facet containing A781 at the end. In this way the target facet benefits from the subtraction of the previous facets, in principle allowing us to achieve nearly thermal noise limited images. \\
\end{enumerate}

\begin{figure}
 \centering
 \includegraphics[width=\hsize,trim={0cm 1.6cm 0cm 0.2cm},clip]{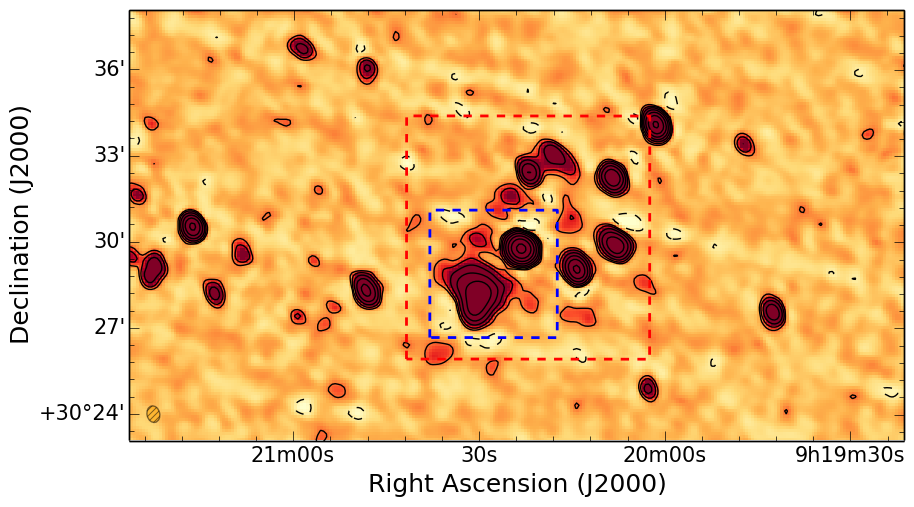}
 \includegraphics[width=\hsize,trim={0cm 0cm 0cm 0.2cm},clip]{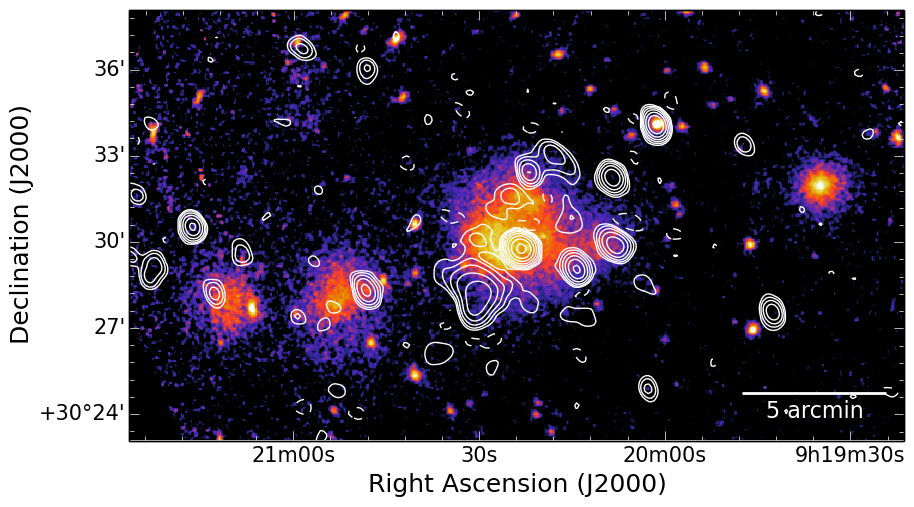}
 \caption{\textit{Top}: \lofar\ 143~MHz image at a resolution of $34.9\arcsec \times 26.6\arcsec$ (the beam is shown in the bottom left corner). Contours are spaced by a factor of 2 starting from $3\sigma$, where $\sigma = 650$ \mujyb. The negative $-3\sigma$ contours are shown in dashed lines. Dashed boxes denote the FOV of the other images reported along the paper. \textit{Bottom}: \xmm\ smoothed image with the \lofar\ contours overlaid.}
 \label{fig:lofar_xmm}
\end{figure}

The \lofar\ images reported in the paper were produced with \wsclean\ v2.4 \citep{offringa14} and  have a central observing frequency of 143~MHz. The imaging was carried out using the multi-scale multi-frequency deconvolution algorithm described in \citet{offringa17}. Data were calibrated (and subsequently imaged) applying an inner \uv-cut of $200\lambda$ to eliminate the noise from the shortest baselines. The largest angular scale that is possible to recover with this \uv-cut is $17.2\arcmin$, larger than the separation between each cluster of the chain. The \uv-tapering of visibilities and the Briggs weighting scheme \citep{briggs95} with different \texttt{robust} values were used to obtain two images with different resolutions. The low-resolution image of the cluster chain is shown in Fig.~\ref{fig:lofar_xmm}. \\
\indent
It is known that the \lofar\ flux density scale can show systematic offsets and needs to be corrected relying on other surveys \citep[\eg][]{vanweeren16calibration, hardcastle16}. In this respect, we cross-matched a catalogue of \lofar\ point sources extracted in the facet containing A781 with the \wenssE\ at 325~MHz \citep[\wenss;][]{rengelink97}. We rescaled the \wenss\ flux densities at 143~MHz assuming a spectral index $\alpha=0.75$. The adopted correction factor of 0.85 on \lofar\ flux densities was derived from the mean flux density ratio \lofar/\wenss$_{143}$. We conservatively set a systematic uncertainty of 20\% on \lofar\ flux density measurements.

\subsection{\gmrt}

\citet{venturi08, venturi11} presented \gmrt\ observations of A781 at 325~MHz and at 610~MHz. In this work, we reanalysed these data sets with the \spamE\ (\spam) package \citep{intema09} and produced new images of the cluster. The details of the observations are shown in Tab.~\ref{tab:radio_obs}. The data reduction with \spam\ consists of a standard-automated pipeline that includes data averaging, instrumental calibration, multiple cycles of self-calibration, and flagging of bad data. Furthermore, the bright sources within the primary beam are selected and used to perform a direction-dependent calibration, whose solutions are interpolated to build a global ionospheric model to suppress ionospheric phase errors. The calibrated data are then reimaged with \wsclean\ v2.4 \citep{offringa14}, as described at the end of Section~\ref{sec:lofar}.  Further details on the \spam\ pipeline are provided in \citet{intema09, intema17}. The flux density scale in the images was set by calibration on 3C48 (at 610 MHz) and 3C286 (at 325 MHz) using the models from \citet{scaife12}. No flux scale offset (\eg\ from the system temperature; see \citealt{sirothia09}) was found by cross-matching a catalogue of \gmrt\ sources with the \wenss\ \citep{rengelink97}. Residual amplitude errors are estimated to be 15\% at 325~MHz and 10\% at 610~MHz, which agrees with other studies \citep[\eg][]{chandra04}.

\subsection{XMM-Newton}

The Abell 781 complex was observed twice with \xmm\ (\obsid: 0150620201 and 0401170101), for a total exposure time of 98.7 ks. Data reduction was performed using the pipeline developed to analyse the observations of the \xcopE\ \citep[\xcop;][]{eckert17xcop}, which is fully described in \citet{ghirardini18arx}. The pipeline uses the \esasE\ (\esas) developed within the \xmm\ \sasE\ (\sas\ v14.0.0) to analyse \epicE\ (\epic) observations. Briefly, the tasks \texttt{mos-filter} and \texttt{pn-filter} were used to filter out observation periods affected by soft proton flares. Residual soft proton flare contamination was checked by measuring in a hard band the count rates of the MOS and pn cameras in the exposed and unexposed parts of the detectors FOV (inFOV/outFOV; see \citealt{leccardi08}). The results of this procedure are summarised in Tab.~\ref{tab:xmm_obs}. For MOS cameras, values of inFOV/outFOV below 1.15 indicate absence of residual soft proton flares while values between 1.15 and 1.30 indicate a slight contamination of soft proton flares. Single detector count images were then combined to produce the mosaic \epic\ background-subtracted and exposure-corrected images in the $0.5-2.0$ \kev\ band shown in Figs.~\ref{fig:smoothed} and \ref{fig:lofar_xmm}. The tasks \texttt{ewavelet} and \texttt{cheese} were used to detect and exclude point sources before the spectral region extraction and surface brightness profile fitting. The output files of the routines were checked for missed sources and/or false detections; therefore contaminating point sources were excised. \\
\indent
Spectra of the two \obsid s were extracted in the same regions and jointly fitted in the $0.5-12.0$ \kev\ band (MOS detectors) and in the $0.5-14.0$ \kev\ band (pn detector) with \xspec\ v12.9.0o \citep{arnaud96} adopting Cash statistics \citep{cash79}. The energy range $1.2-1.9$ \kev\ was excluded in the fit owing to strong instrumental emission lines; for the pn detector, we also excluded the range $7.0-9.2$ \kev\  for the same reason. The non-X-ray background was modelled with a phenomenological model that includes a number of fluorescence lines \citep[see][]{ghirardini18arx}. The local sky background was estimated in a cluster free region adopting a model composed of a cosmic X-ray background component, which was modelled with an absorbed power law with photon index $\Gamma = 1.46$ \citep{deluca04}, and of a Galactic foreground component, which was modelled with two thermal plasmas (one unabsorbed and the other absorbed) with solar metallicity and temperatures $0.11$ \kev\ and $0.28$ \kev\ \citep{mccammon02}. The ICM emission was modelled with an absorbed thermal model with normalization, metallicity, and temperature free to vary in the fit. Galactic absorption in the direction of the cluster was set to $\nh = 1.65 \times 10^{20}$ cm$^{-2}$ \citep{kalberla05}. \\
\indent
Surface brightness profiles were extracted and fitted with \proffit\ v1.5 \citep{eckert11} from the \epic\ mosaic image in the $0.5-2.0$ \kev\ band. All the profiles reported in the paper were convolved for the \xmm\ point spread function that was modelled with the \texttt{psf} task \citep[for more details, see Appendix C in][]{eckert16xxl}.

\begin{table}
 \centering
 \caption{Clean exposure time and inFOV/outFOV ratio of each \epic\ detector for the two \xmm\ observations used in this work (medium filter, full frame science mode).}
 \label{tab:xmm_obs}
  \begin{tabular}{lcc} 
  \hline
  & Exposure & inFOV/outFOV \\
  & (ks) & \\
  \hline
  \obsid\ & \multicolumn{2}{c}{0150620201} \\
  \hline
  MOS1 & 14.5 & $1.157\pm0.051$ \\
  MOS2 & 14.2 & $1.070\pm0.044$ \\
  pn & 10.5 & $1.199\pm0.055$ \\
  \\
  \hline
  \obsid\ & \multicolumn{2}{c}{0401170101} \\
  \hline
  MOS1 & 58.6 & $1.106\pm0.025$ \\
  MOS2 & 60.8 & $1.049\pm0.022$ \\
  pn & 47.5 & $1.232\pm0.026$ \\
  \hline
  \end{tabular}
\end{table}

\begin{figure}
 \centering
 \includegraphics[width=.9\hsize]{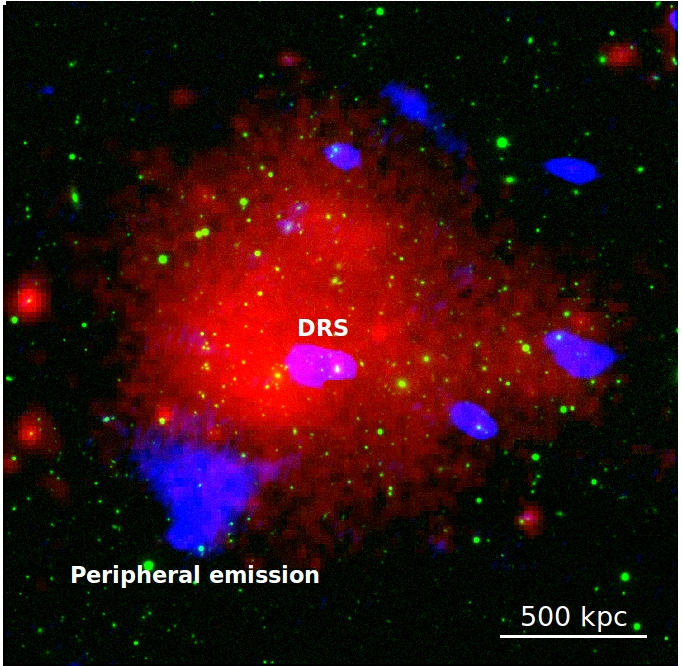}
 \caption{Composite multiwavelength image of A781 (red region of Fig.~\ref{fig:lofar_xmm}). Optical \sdss g,r,i mosaic is shown in green. Radio emission at 143~MHz from \lofar\ is shown in blue. X-ray \xmm\ emission is shown in red.}
 \label{fig:rgb}
\end{figure}

\section{Results}

\begin{figure*}
 \centering
 \includegraphics[width=.33\textwidth,trim={0cm 0cm 0cm 2.6cm},clip]{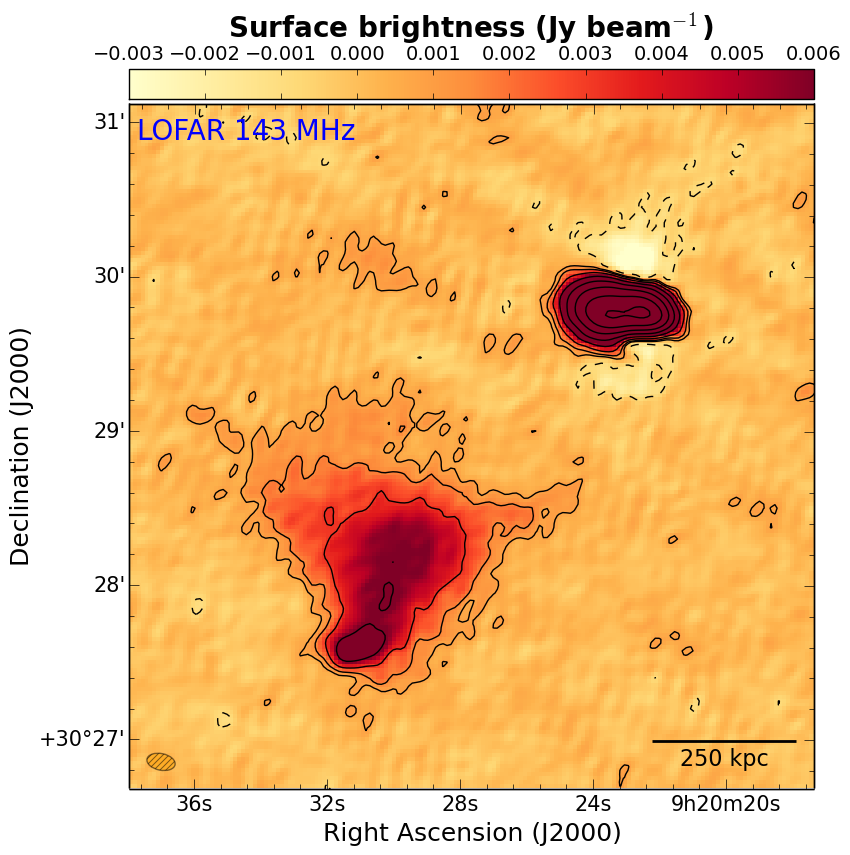}
 \includegraphics[width=.33\textwidth,trim={0cm 0cm 0cm 2.6cm},clip]{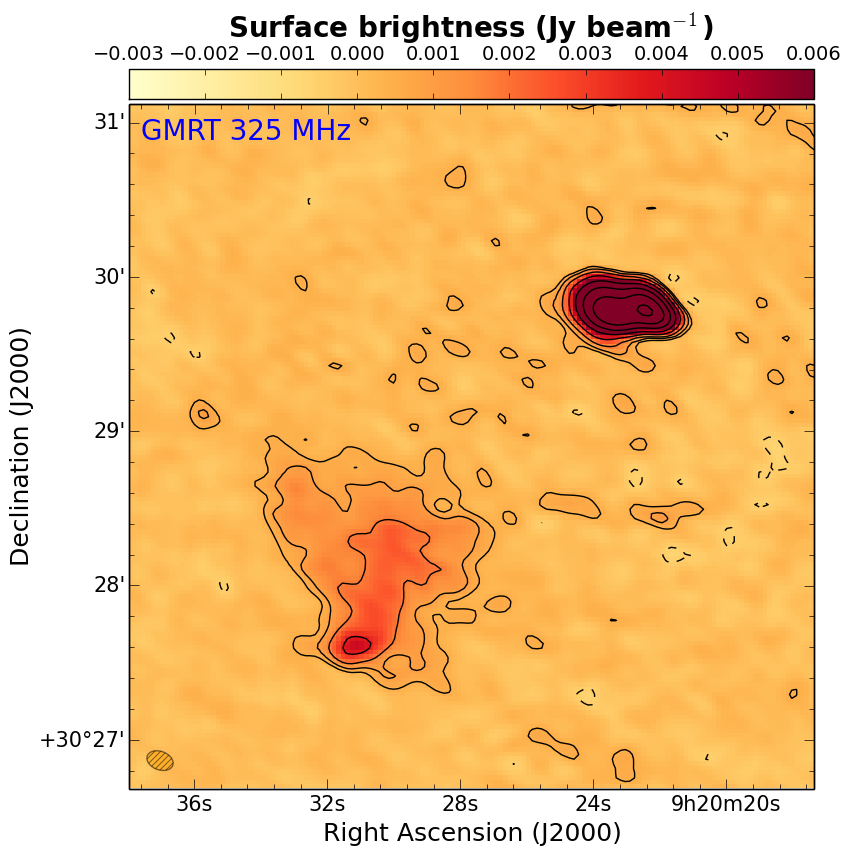}
 \includegraphics[width=.33\textwidth,trim={0cm 0cm 0cm 2.6cm},clip]{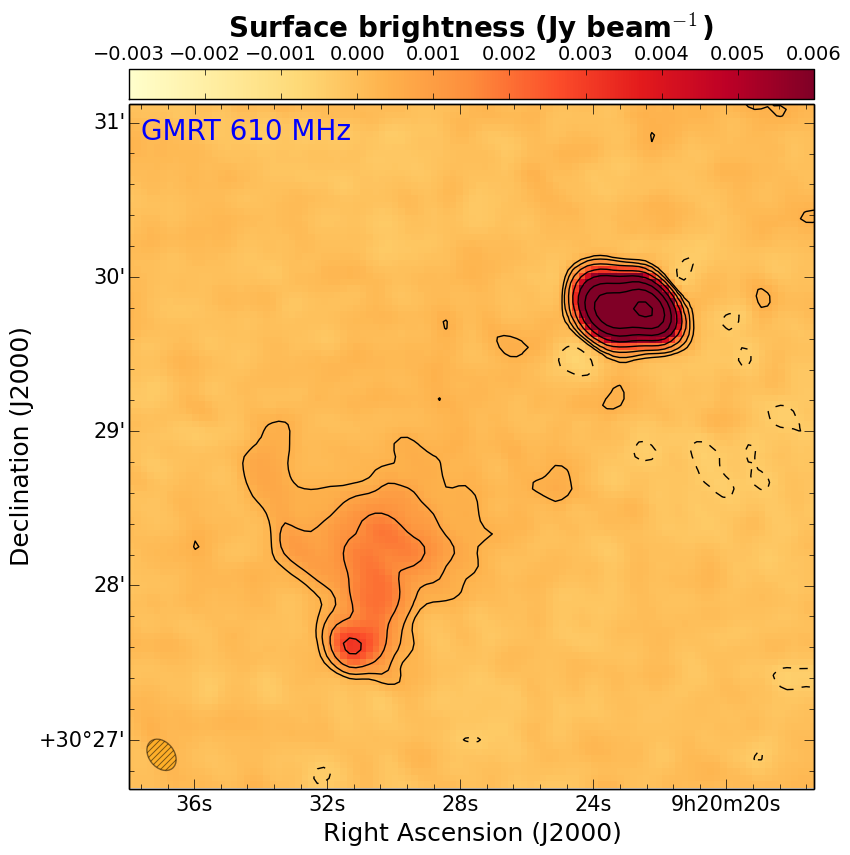}
 \caption{Peripheral emission in A781 (blue region of Fig.~\ref{fig:lofar_xmm}) as observed in the radio band with \lofar\ at 143~MHz (\textit{left}) and with the \gmrt\ at 325~MHz (\textit{centre}) and 610~MHz (\textit{right}). Contours are spaced by a factor of 2 starting from $3\sigma$, where $\sigma_{\rm 143} = 270$ \mujyb, $\sigma_{\rm 325} = 150$ \mujyb\ , and $\sigma_{\rm 610} = 120$ \mujyb. The negative $-3\sigma$ contours are shown in dashed. The beam sizes are $11.1\arcsec \times 6.5\arcsec$ (143~MHz), $10.6\arcsec \times 7.2\arcsec$ (325~MHz) and $13.5\arcsec \times 9.8\arcsec$ (610~MHz) and are shown in the bottom left corners.}
 \label{fig:radio_images}
\end{figure*}

\subsection{The peripheral emission in A781}

\begin{figure}
 \centering
 \includegraphics[width=\hsize,trim={0cm 7.0cm 0cm 0.5cm},clip]{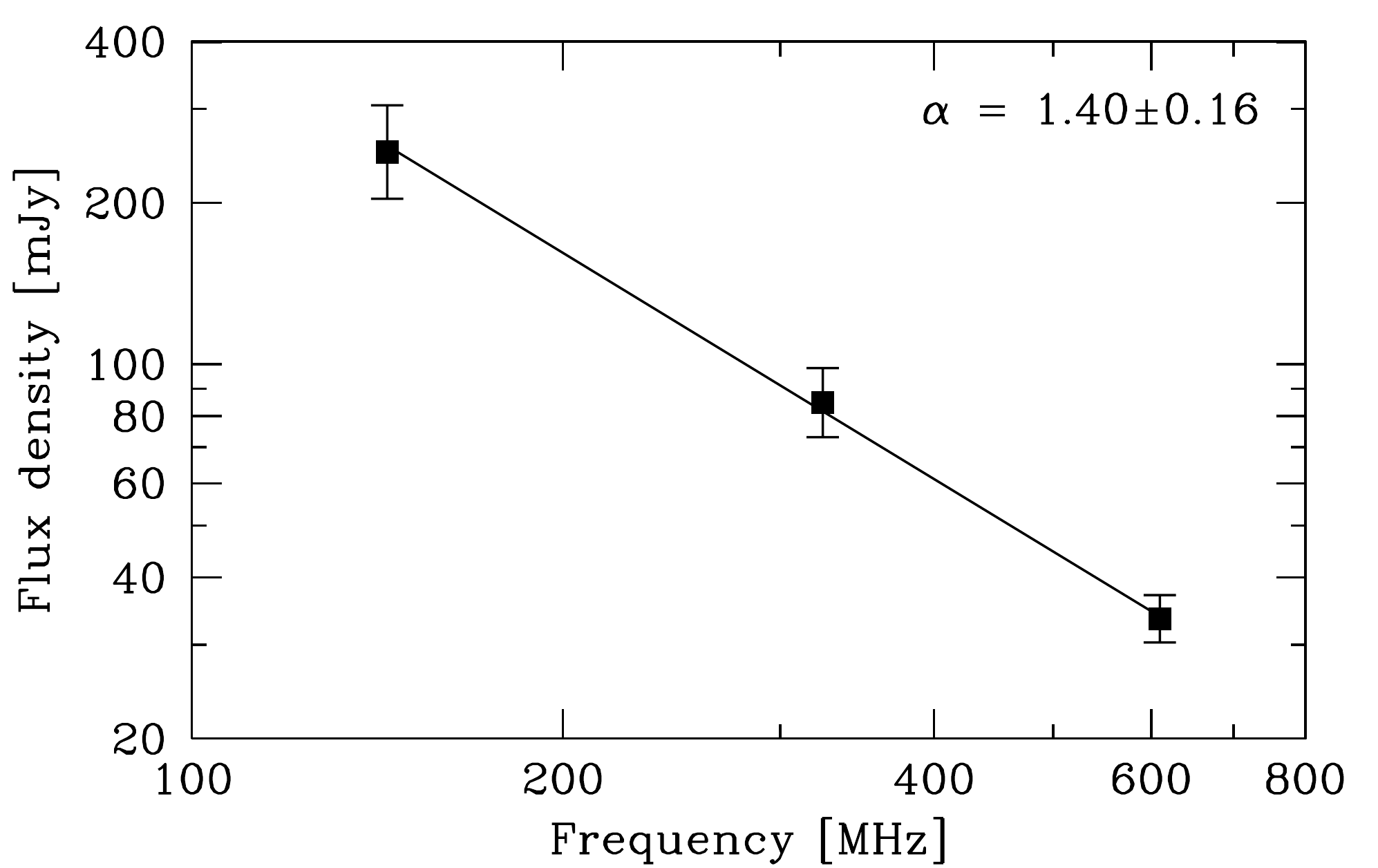}
 \caption{Integrated spectrum of the peripheral source in A781.}
 \label{fig:spectrum}
\end{figure}

\begin{figure}
 \centering
 \includegraphics[width=\hsize]{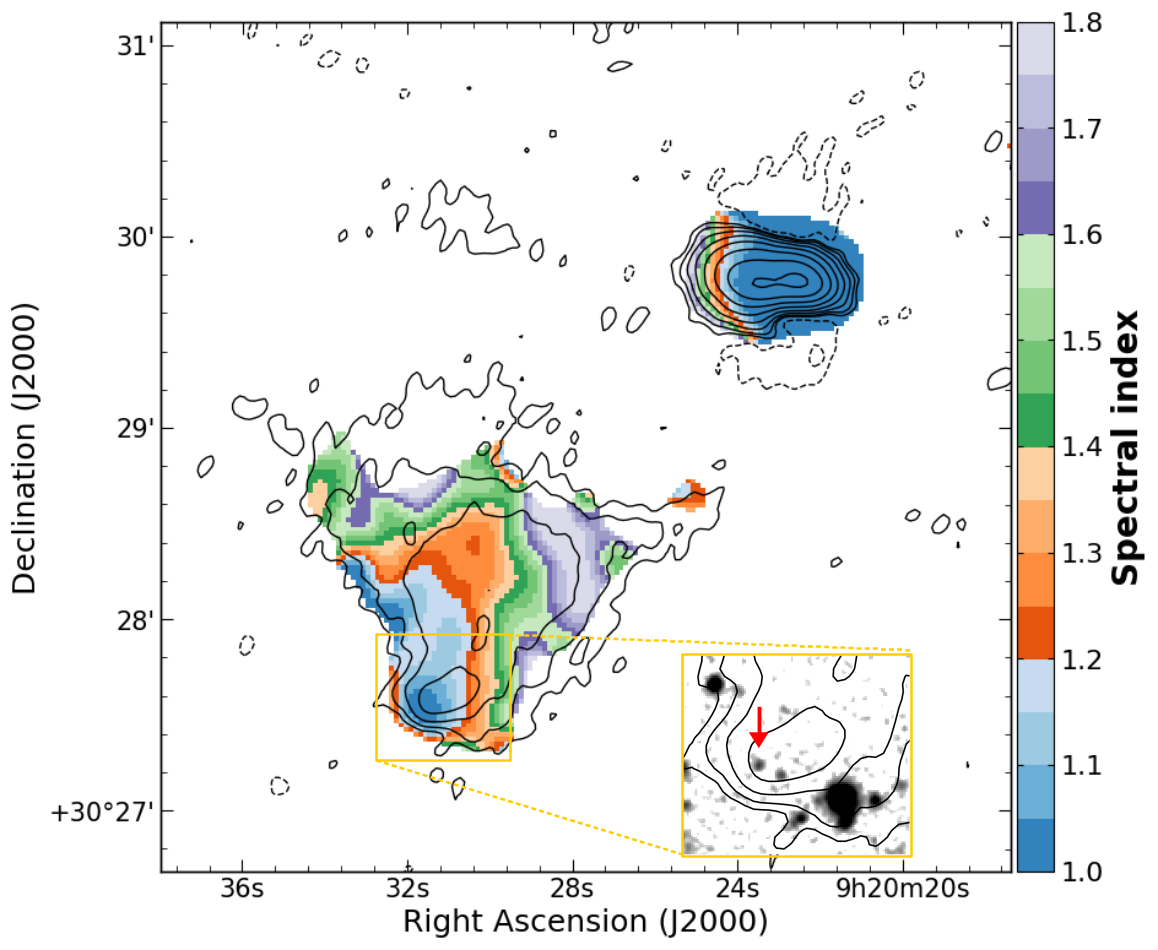}
 \caption{Spectral index map of the peripheral emission between 143~MHz and 610~MHz at a resolution of $15\arcsec \times 15\arcsec$ overlaid on the \lofar\ contours of Fig.~\ref{fig:radio_images}. Pixels with values below $3\sigma$ were blank. The corresponding error map is reported in Fig.~\ref{fig:alpha_error}. The inset panel shows an \sdss\ image with two candidate optical counterparts.}
 \label{fig:alpha_map}
\end{figure}

\begin{figure}
 \centering
\includegraphics[width=\hsize,trim={0cm 10.8cm 0cm 0cm},clip]{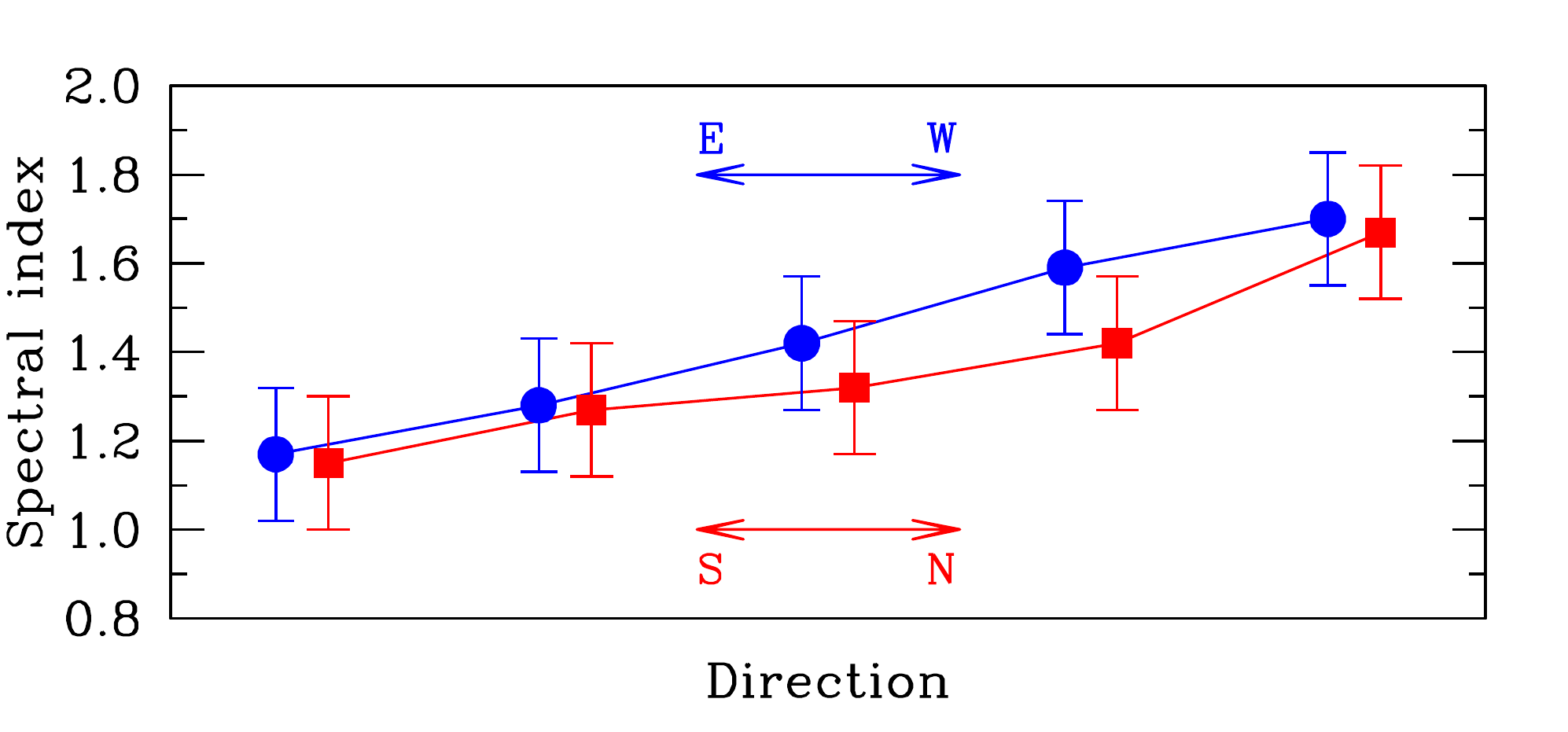} 
 \includegraphics[width=.45\hsize]{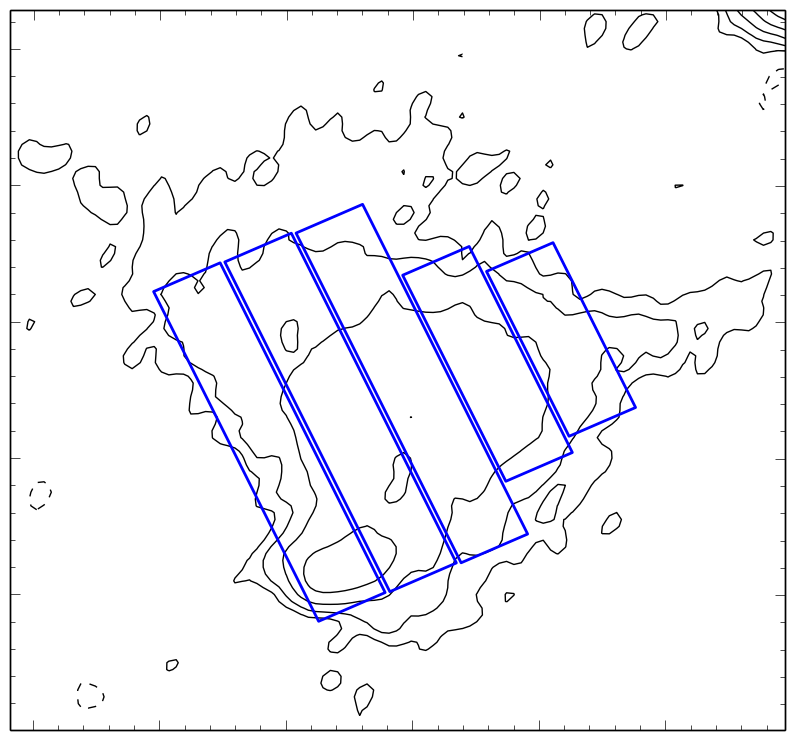}
 \includegraphics[width=.45\hsize]{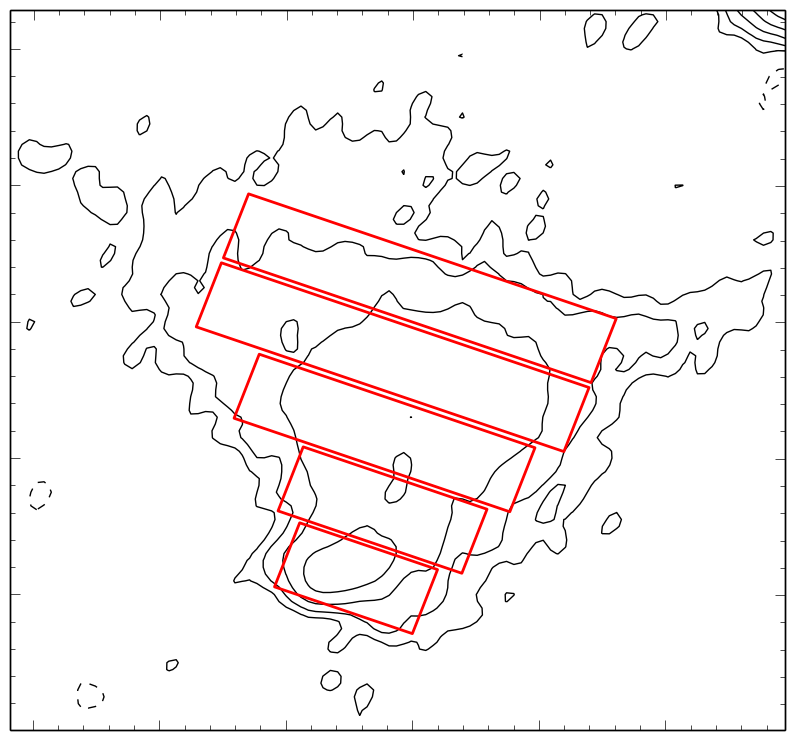}
 \caption{Spectral index gradient of the peripheral emission towards the E-W (blue) and S-N (red) directions. The spectral index has been computed between 143 and 610 MHz in the sectors shown in the bottom panels.}
 \label{fig:gradient}
\end{figure}

The peripheral diffuse radio emission in the SE outskirts of the main cluster (Fig.~\ref{fig:rgb}) was classified as a candidate radio relic by \citet{venturi08}. The source has been observed with the \gmrt\ \citep{venturi08, venturi11, venturi13} and \vla\ \citepalias{govoni11}; it is also detected in the \nvssE\ \citep[\nvss;][]{condon98} but not in the \firstE\ survey \citep[\first;][]{becker95}. \\
\indent
In Fig.~\ref{fig:radio_images}, we show our images of the peripheral emission at three frequencies with comparable resolution obtained from the new \lofar\ data and from the reanalysis of the archival \gmrt\ observations. The flux densities measured within the \lofar\ $3\sigma$ contour in these images are $S_{\rm 143\,MHz} = 267 \pm 53$ m\jy, $S_{\rm 325\,MHz} = 94 \pm 14$ m\jy,\ and $S_{\rm 610\,MHz} = 38 \pm 4$ m\jy, where the quoted errors are given by the uncertainty in the flux density scale and the noise of the images weighted for the number of beams added in quadrature. The source morphology is consistent between 143~MHz and 610~MHz, appears slightly more extended at low frequency, and has a largest linear size of $\sim550$ kpc. The source displays a peculiar wedge shape characterised by a bright knot of emission in the SE that is attached to a high surface brightness spine that is extended NW in the direction of the central double radio source (DRS; \cf\ Fig.~\ref{fig:rgb}). The radio emission shows a sharper edge towards the E direction where the X-ray thermal emission also fades away. \\
\indent
We measured the spectral index properties of the source from images produced with a \texttt{uniform} weighting scheme and with matched \uv-range. The integrated spectral index computed between the three frequencies is $\alpha=1.40\pm0.16$ (Fig.~\ref{fig:spectrum}), which is consistent within the errors with that reported by \citetalias{venturi11}. The $k$-corrected and spectral index rescaled radio power of the source at 1.4~GHz is $P_{1.4\,\rm{GHz}} \sim 3.5 \times 10^{24}$ \whz, assuming that it is located at the cluster redshift $z=0.3004$. The spectral index map calculated from the 143~MHz and 610~MHz images convolved to the same resolution of $15\arcsec \times 15\arcsec$, corrected for any position misalignment, and regridded to identical pixel size, is shown in Fig.~\ref{fig:alpha_map} (the error map is reported in Appendix~\ref{app:erro}). This map shows that the SE bright knot of emission also has a flatter spectral index, possibly arising from the radio emission of an active galactic nucleus (AGN). In the \sdssE\ \citep[\sdss;][]{york00}, two possible optical counterparts are observed in this position (see inset panel in Fig.~\ref{fig:alpha_map}); these are discussed in Section~\ref{sec:nature}. The absence of significant compact emission at a level of 0.5 m\jy\ beam$^{-1}$ in the \first\ data suggests that the AGN is not active anymore. The diffuse source exhibits a hint of spectral index flattening in coincidence with the E edge of the radio emission. The spectral index gradually steepens in the direction of the DRS, where $\alpha\sim1.8$. A similar spectral trend can be inferred also from Fig.~5 of \citetalias{govoni11}, despite the lower resolution ($53\arcsec \times 53\arcsec$) of their spectral index map. As a further check, we evaluated the spectral index of the peripheral source in sectors, as shown in Fig.~\ref{fig:gradient}. This confirms a spectral gradient in both the E-W and S-N directions. We mention that spectral index steepening towards the cluster centre has been observed in a number of radio relics \citep[\eg][]{giacintucci08, vanweeren10, degasperin15double, hoang18a1240}.

\subsection{X-ray discontinuities in the ICM}

\begin{figure}
 \centering
 \includegraphics[width=\hsize]{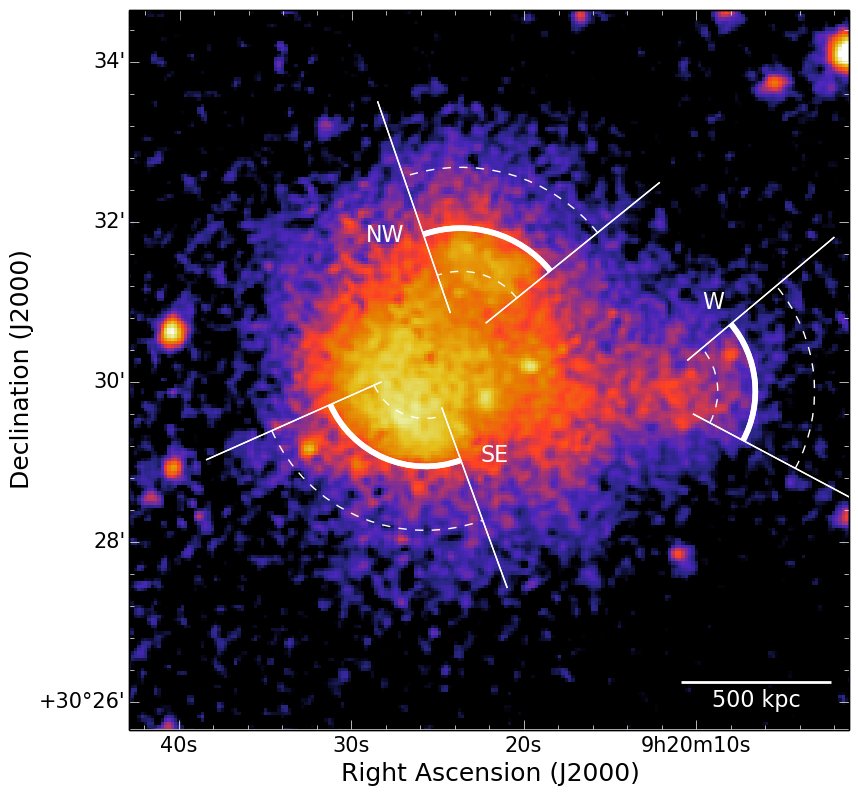}
 \caption{\xmm\ smoothed image of A781 (red region of Fig.~\ref{fig:lofar_xmm}) with the sectors used for the spectral and spatial analysis. Thick lines denote the position of the edges; dashed lines limit the regions used for the spectral analysis.}
 \label{fig:sectors}
\end{figure}

\begin{figure*}
 \centering
 \includegraphics[width=.33\textwidth,trim={0.5cm 0.2cm 1.5cm 0.5cm},clip]{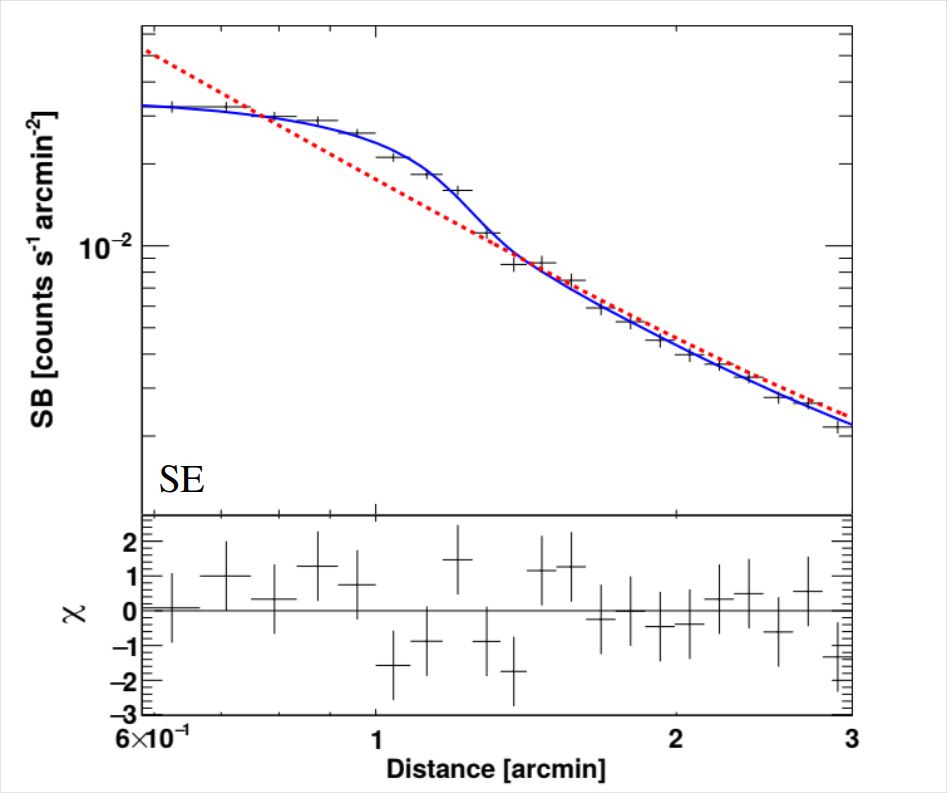} 
 \includegraphics[width=.33\textwidth,trim={0.5cm 0.2cm 1.5cm 0.5cm},clip]{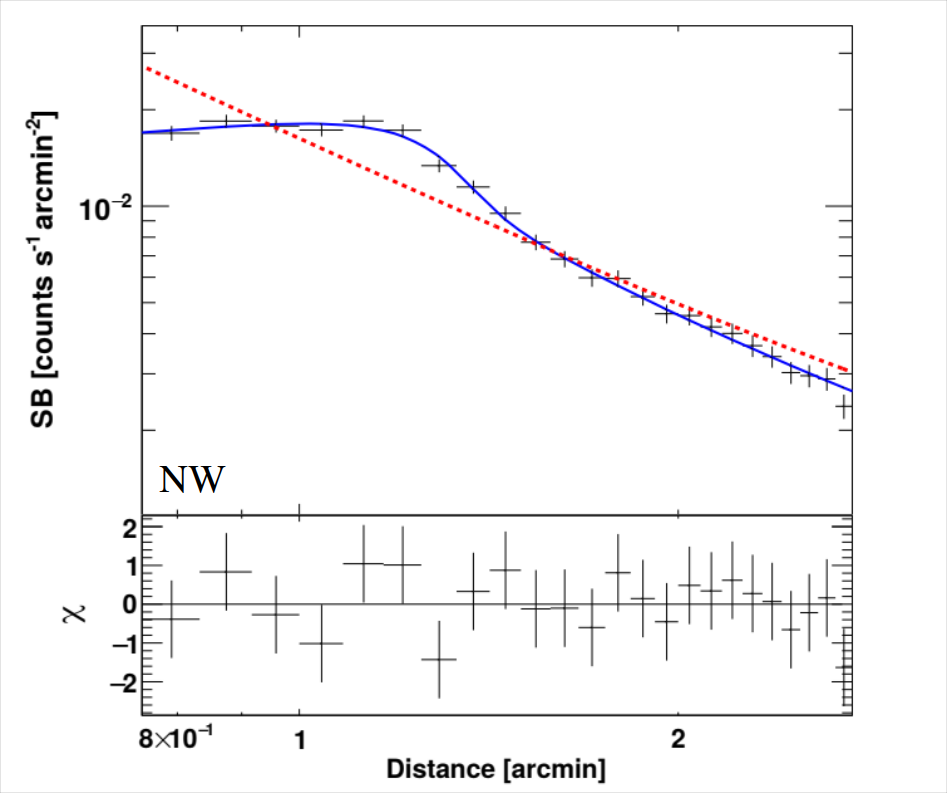}
 \includegraphics[width=.33\textwidth,trim={0.5cm 0.2cm 1.5cm 0.5cm},clip]{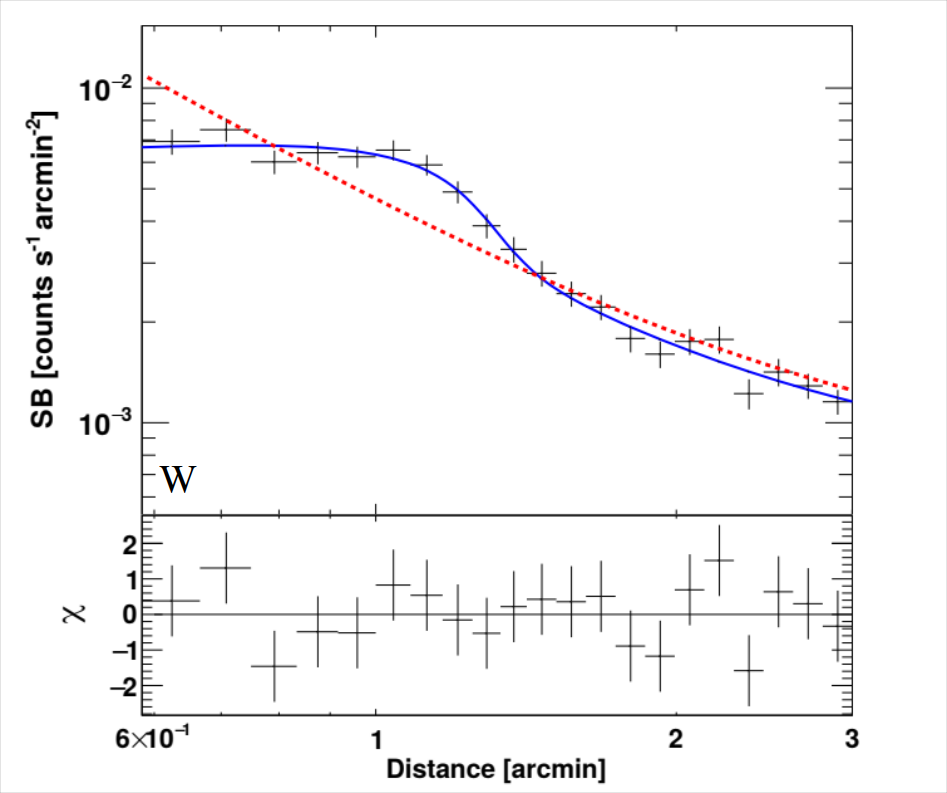}
 \caption{\xmm\ surface brightness profiles in the $0.5-2.0$ \kev\ energy band extracted in the white sectors of Fig.~\ref{fig:sectors}. The best-fitting broken power laws with residuals and single power laws are reported in solid blue and dashed red, respectively. The residuals at the bottom of the plots refer to the broken power-law fits.}
 \label{fig:edges}
\end{figure*}

The visual inspection of the \xmm\ image in the $0.5-2.0$ \kev\ band suggests the presence of three surface brightness jumps in A781, towards the SE, NW, and W directions, that have not been studied in the literature so far. We investigated the possible features with the fitting of the surface brightness profiles extracted in the sectors highlighted in Fig.~\ref{fig:sectors}. A broken power-law model was assumed to fit the data as it generally provides a good description of discontinuities in the ICM, namely shocks and cold fronts \citep[\eg][]{markevitch07rev, owers09sample, botteon18edges}. A single power-law model was also fitted for comparison. The three profiles are shown in Fig.~\ref{fig:edges}. The broken power-law models always yield the best description of the data, confirming the existence of drops in surface brightness. The compression ratios between the downstream and upstream density are $\compr = 1.9 \pm 0.1$ (SE), $\compr = 2.0 \pm 0.2$ (NW), and $\compr = 2.2^{+0.4}_{-0.3}$ (W). 

\begin{table}
 \centering
 \caption{Properties measured across the X-ray surface brightness discontinuities.}
 \label{tab:temperatures}
  \begin{tabular}{lcccc} 
  \hline
   & SE & NW & W \\
  \hline
  \ktd\ (\kev) & $5.4^{+0.4}_{-0.2}$ & $4.1^{+0.2}_{-0.2}$ & $4.2^{+0.6}_{-0.4}$ \\
  \ktu\ (\kev) & $9.5^{+1.6}_{-1.3}$ & $7.4^{+1.0}_{-1.0}$ & $2.6^{+0.5}_{-0.5}$ \\
  \compr\ & $1.9^{+0.1}_{-0.1}$ & $2.0^{+0.2}_{-0.2}$ & $2.2^{+0.4}_{-0.3}$ \\
  $\mathcal{P}$ & $1.0^{+0.1}_{-0.1}$ & $1.1^{+0.1}_{-0.1}$ & $3.5^{+0.9}_{-0.8}$ \\
  \hline
  \end{tabular}
\end{table}

In order to determine the nature of the edges (shocks or cold fronts), a careful spectral analysis is necessary. Shocks are characterised by higher temperature and pressure in the downstream region than in the upstream region. Instead, the temperature jump is inverted and the pressure is almost continuous across cold fronts. We extracted and fitted spectra in the downstream and upstream regions delimited by the dashed and solid lines in Fig.~\ref{fig:sectors}. The pressure jump $\mathcal{P}$ at the discontinuity can be computed as the product between the density and temperature ratios\footnote{Although this procedure combines a deprojected density jump with a temperature evaluated along the line of sight, previous studies have shown that projection effects do not have a strong impact \citep[\eg][]{botteon18edges}.} achieved with the spatial and spectral analysis, respectively. Results are summarised in Tab.~\ref{tab:temperatures}. All the surface brightness discontinuities are associated with temperature jumps. For the SE and NW edges, the downstream temperature is lower and the pressure is consistent to be constant across the discontinuities, as expected in the case of cold fronts. For the W edge, the downstream gas is hotter and a pressure jump is observed, revealing the shock nature of the discontinuity. We applied the Rankine-Hugoniot equations \citep[\eg][]{landau59} to derive independent constraints of the shock Mach number from the temperature and density jumps, leading to consistent values of $\machkt=1.6\pm0.3$ and $\machsb=1.9^{+0.4}_{-0.3}$, respectively. 

\begin{figure}
 \centering
 \includegraphics[width=\hsize]{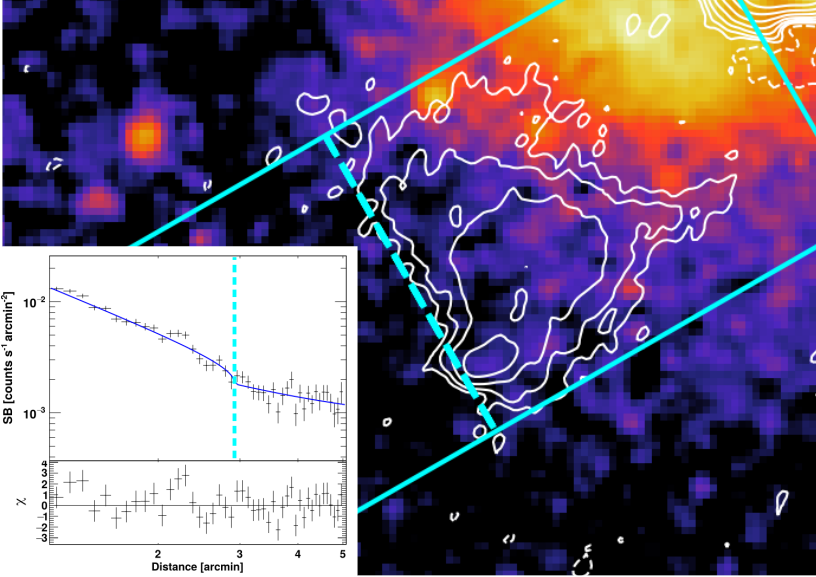}
 \caption{Sector used to extract the \xmm\ surface brightness profile across the peripheral source. In the fit, the position of the jump was fixed at the location of the radio edge of emission (dashed line).}
 \label{fig:box}
\end{figure}

Finally, we searched for a possible X-ray discontinuity at the position of the peripheral diffuse radio emission. In particular, a shock could be responsible for the peculiar morphology and the observed spectral index trend of the source (Fig.~\ref{fig:alpha_map}). Moreover, a number of merger shocks have been found ahead of cold fronts \citep[\eg][]{vikhlinin01cold, markevitch02bullet, russell10, russell12, emery17}, and it is also possible that the SE cold front detected in A781 follows a shock. In this respect, we extracted and fitted a surface brightness profile in a box across the radio edge in the E that shows a hint of spectral index flattening, as shown in Fig.~\ref{fig:box}. However, the current \xmm\ data is not deep enough to characterise this potential feature as a consequence of the low count statistics of this region. We used the \textsc{MultiNest} Bayesian nested sampling algorithm \citep{feroz09} interfaced in \proffit\ to determine an upper limit of $\compr <1.6$ (90\% confidence level) on the compression factor by fitting a broken power law and assuming that the discontinuity is locate at the edge of the radio emission. This implies that if a shock exists, it is weak ($\mach < 1.4$). Projection effects (if any) should play a small role as the detection of the two diametrically opposite cold fronts in the NW and SE directions suggests that the merger occurs approximately on the plane of the sky.

\subsection{Constraints on the radio halo emission}

The presence of diffuse radio emission at the centre of A781 was previously uncertain. Originally, A781 was classified as one of the few dynamically disturbed without evidence of a radio halo in the \gmrt\ 610~MHz sample of \citet{venturi08}. However, \citetalias{govoni11} claimed the presence of a radio halo using \vla\ observations and recorded a flux density $S_{\rm 1.4\,GHz} = 20\pm5$ mJy at 1.4~GHz. Nonetheless, this detection remained uncertain as \citetalias{venturi11} found only a low level of residuals in the cluster centre with the \gmrt\ at 325~MHz (consistent with our reanalysis of the same data set performed with the \spam\ pipeline), which would imply an unusual flat spectrum $\alpha^{\rm 1.4\,GHz}_{\rm 325\,MHz} < 0.5$ for a diffuse cluster source when combined with the claim of \citetalias{govoni11}. The \lofar\ has the sensitivity required to shed light on this point: our images have a brightness sensitivity $1.5-2.5$ times better than the \gmrt\ at 325~MHz and the \vla\ at 1.4~GHz assuming a typical value of $\alpha=1.3$ for the radio halo spectrum\footnote{This estimate is also conservative as it does not account for the fact that the \uv-coverage at short baselines of \lofar\ is much better than that of the \gmrt\ and \vla.}. With this spectral index and considering the flux density reported with the \vla\ by \citetalias{govoni11}, the flux density expected at 143~MHz is $\sim400$ m\jy. This should be clearly observable in the \lofar\ image. However, a radio halo is not visible and only a low level of residuals is measured in the central region of A781 at 143 MHz. The origin of these residuals is unclear. They may well be patches of emission due to unresolved sources in the cluster or possible contamination of spurious emission due to the central bright DRS ($S_{\rm 143\,MHz} = 0.5\pm0.1$ \jy).

\begin{table}
 \centering
 \caption{Expected properties of a radio halo in A781 according to the relation of \citet{cassano13}. The halo reference radius was calculated as $r_h = 2.6r_e$ \citep{bonafede17}.}
 \label{tab:halo_cassano}
  \begin{tabular}{ccccc} 
  \hline
   \mfive\ & $P_{1.4\,\rm{GHz}}$ & $S_{1.4\,\rm{GHz}}$ & $r_h$ & $r_e$ \\
   (\msun) & (\whz) & (m\jy) & (kpc) & (kpc) \\
  \hline
   $6.1 \times 10^{14}$ & $1.56 \times 10^{24}$ & 5.0 & 437 & 168 \\
  \hline
  \end{tabular}
\end{table}

To further quantify the limits of our non-detection, we used the technique of injecting mock radio haloes in the data set to infer an upper limit on the diffuse emission flux density \citep[\eg][]{brunetti07cr, venturi08, kale13, kale15, bonafede17, cuciti18}. Specifically, we applied this method to A781 following the procedure described in \citet{bonafede17}. The mock haloes were injected in a region close to the cluster centre, avoiding bright radio sources and selecting a region with similar noise properties to that within the cluster region. The surface brightness of the mock radio haloes is assumed to follow an exponential law in the form $I(r)=I_0 \exp(-r/r_e)$, where $I_0$ is the central surface brightness and $r_e$ denotes the $e$-folding radius \citep[\eg][]{orru07, murgia09}.  We first injected a halo with the properties reported in Tab.~\ref{tab:halo_cassano}, \ie\ consistent with that expected from the $P_{1.4\,\rm{GHz}}-\mfive$ relation of \citet{cassano13} starting from the value of \mfive\ reported in the PSZ2 catalogue \citep{planck16xxvii}. We verified that this mock radio halo was clearly detected by our \lofar\ observation at 143 MHz (assuming a spectral index $\alpha=1.3$, the implied flux density is $S_{143\,\rm{MHz}} = 97$ m\jy). We then reduced the flux density of the injected haloes until we recovered a flux density that matches the level of residuals measured in the cluster centre. This occurred when $S_{143\,\rm{MHz}} < 50$ m\jy, and we consider this as the upper limit on the radio halo emission. This converts into a limit of $P_{143\,\rm{MHz}} < 1.6 \times 10^{25}$ \whz\ for the radio halo power at 143~MHz. Whilst the \lofar\ brightness sensitivity is much better than that of the \gmrt, this limit is similar to that derived by \citet{venturi08}. Indeed, the residuals due to the contamination from the DRS constrain the depth of our measurement. The upper limit is a factor of 2 below the values expected by the $P_{1.4\,\rm{GHz}}-\mfive$ relation. We note that the \planck\ estimate of \mfive\ for A781 could be slightly biased high because of the presence of the ``Middle'' cluster in the \planck\ beam \citep[see][for a similar case]{botteon18a1758}. \\
\indent
There is evidence that a fraction of merging clusters do not show radio haloes and this fraction is seen to increase at smaller cluster masses \citep{cuciti15}. According to current models, a fraction of these low-mass merging clusters should glow at low radio frequencies and host haloes with very steep spectra \citep[\eg][]{cassano06, brunetti08} that are also typically less luminous in the $P_{1.4\,\rm{GHz}}-\mfive$ plane than radio haloes with flatter spectrum \citep[\eg][]{cassano10mc, wilber18a1132}. Unfortunately, the artefacts around the DRS prevent us from exploring the presence of a halo less luminous than a halo in line with the \citet{cassano13} relation. \\
\indent
We also searched our low-resolution \lofar\ image for emission from the other clusters in the Abell 781 chain (Fig.~\ref{fig:lofar_xmm}). There are no clear detections towards any of the other clusters but this is to be expected given the low mass of these components \citep[\cf\ Tab.~1 in][]{wittman14}. Owing to the expected non-detections we did not determine precise upper limits on the diffuse radio emission.

\section{Discussion}

\subsection{Nature of peripheral radio emission}\label{sec:nature}

\begin{figure*}
 \centering
 \includegraphics[width=.45\textwidth,valign=c]{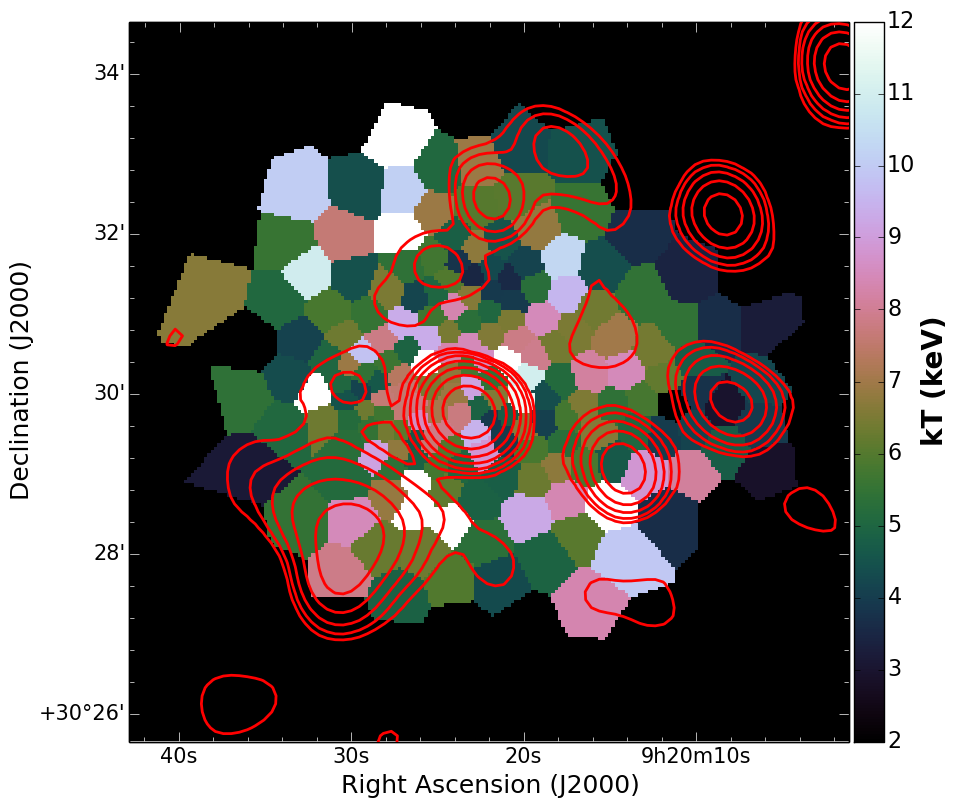}
 \includegraphics[width=.45\textwidth,valign=c]{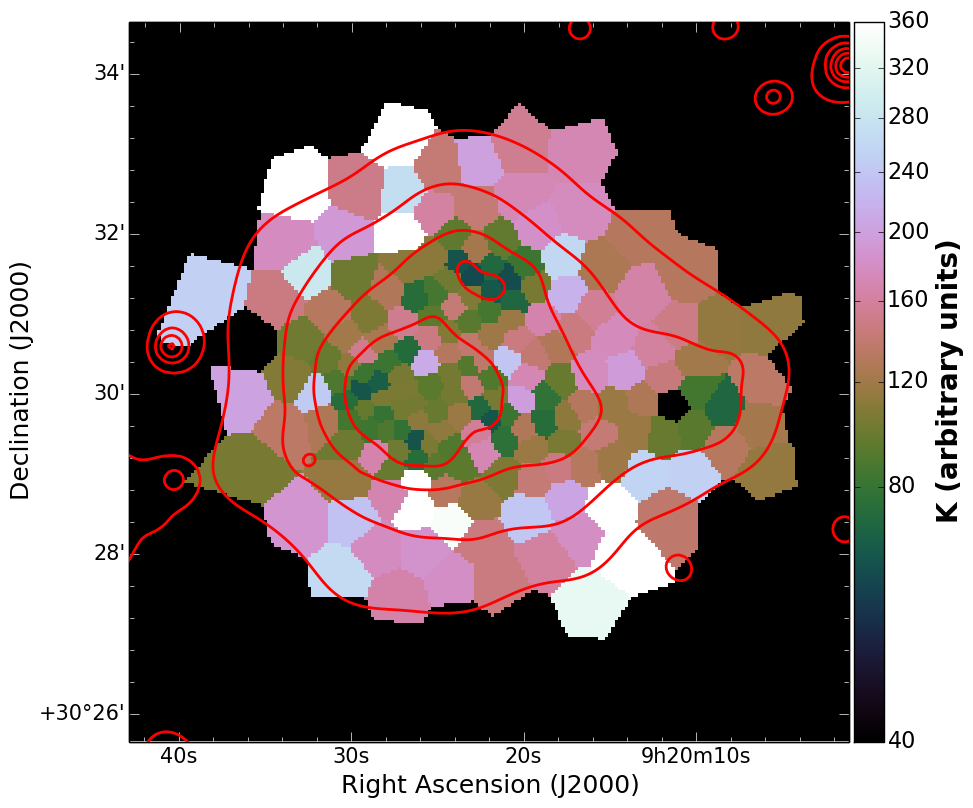}
 \caption{Thermodynamical properties of the ICM in A781 projected along the line of sight. \textit{Left}: temperature map with overlaid the \lofar\ contours of Fig.~\ref{fig:lofar_xmm}. \textit{Right}: entropy map with overlaid the \xmm\ contours of Fig.~\ref{fig:smoothed}. The corresponding error maps are reported in Fig.~\ref{fig:thermo_error}.}
 \label{fig:icm_maps}
\end{figure*}

The most striking feature in the composite image of A781 (Fig.~\ref{fig:rgb}) is the peculiar peripheral radio source in the SE. Whilst its nature is still uncertain, two possible explanations for its origin can be proposed based on the results coming from the joint radio and X-ray analysis presented in this work. \\
\indent
The first possibility is that the source traces a radio relic, as already hypothesised by \citet{venturi08}. This scenario agrees with the location of the emission in the cluster outskirts and the overall steepening of the spectral index towards the cluster centre (Fig.~\ref{fig:alpha_map} and \ref{fig:gradient}). Although the source has an edge that coincides with a region with flatter spectral index as observed in almost the totality of radio relics, the global morphology does not recall the typical arc-shaped structure observed for this class of sources \citep[\eg][]{vanweeren10}. In particular, the presence of the high surface brightness spine and the bright knot of emission with flat spectral index are difficult to explain in the radio relic scenario even assuming strong projection effects \citep[\eg][]{slee01, hoeft08}. \\
\indent
Alternatively, the source could be associated with a radio galaxy just turned off (as suggested by the lack of a bright core in the \first). Whilst its morphology does not befit directly to any of the typical classes of radio galaxies \citep[\eg][]{miley80rev}, the structure observed in Fig.~\ref{fig:radio_images} vaguely resembles a head tail source. In this case, it is natural to associate the core emission with the bright knot in the SE that displays a flatter spectral index (Fig.~\ref{fig:alpha_map}). Thus, the high surface brightness spine would result from the relativistic plasma trailed behind the host galaxy during its motion towards the cluster outskirts. As a consequence of particle ageing, the spectral index gets steeper along the tail; however, in A781, the spectral index also shows a transversal trend (Fig.~\ref{fig:gradient}). We tentatively interpret this gradient as the signature of a shock passing through the radio galaxies from the W to the E direction compressing and potentially re-accelerating the radio plasma. The interaction between shocks and radio galaxies is very complicated and leads both to the compression of the plasma and the modification of the source morphology \citep[\eg][]{ensslin02relics, pfrommer11, jones17}. In this scenario, the presence of a clear spectral gradient would suggest that the shock has gone through the tail. An external shock can only propagate as a shock inside the tail if the sound speed inside the relativistic plasma is lower than the shock speed in the external medium. This could be explained by entrainment of thermal plasma in the non-thermal plasma and a small volume filling fraction of the non-thermal plasma. Tailored numerical simulations on the source in A781 will test this scenario. 

\begin{table}
 \centering
 \caption{Photometric redshifts of the source denoted with the red arrow in the inset panel in Fig.~\ref{fig:alpha_map} for two different \sdss\ data releases from \citet[DR10]{ahn14} and \citet[DR14]{abolfathi18}.}
 \label{tab:photoz}
  \begin{tabular}{lcc} 
  \hline
   & RF method & KD-tree method \\
  \hline
  DR10 & $0.292\pm0.126$ & $0.241\pm0.132$ \\
  DR14 & $-$ & $0.467\pm0.124$ \\
  \hline
  \end{tabular}
\end{table}

Both interpretations described above assume that a shock is involved in the formation of the peripheral source. Nonetheless, the present \xmm\ observations allowed us to determine only an upper limit on the density jump across the E region of the source, which would imply a low Mach number shock. We currently prefer the second scenario as it can be more easily reconciled with the source morphology and spectral index properties. Furthermore, two possible optical counterparts are visible in the \sdss\ image within the radio knot. Both the sources are detected by the \spitzer\ satellite, possibly indicating infrared emission from AGNs. However, only the galaxy indicated with the red arrow\footnote{\sdss\ J092031.54+302733.1} in the inset panel in Fig.~\ref{fig:alpha_map} is in the \sdss\ catalogue. Various estimates\footnote{See \citet{csabai07} and \citet{carliles10} for details on the photometric redshift estimation methods.} of the photometric redshift for this object are reported in Tab.~\ref{tab:photoz}. The galaxy is consistent with being a cluster member within $1\sigma$ for \citet{ahn14} and within $1.4\sigma$ for \citet{abolfathi18}. We mention that the apparent discrepancy between the redshifts reported in the two \sdss\ Data Releases might be due to changes in the machine learning technique (\eg\ in the training sample) between the two releases. Spectroscopic follow-up observations are required to precisely determine the galaxy redshift and nuclear activity. \\
\indent
In conclusion, we point out that the radio relic and radio galaxy--shock interaction scenarios do not necessarily exclude each other. The shock with $\mach < 1.4$ inferred from the X-ray analysis, if present, would challenge DSA owing to inefficient particle acceleration at weak cluster shocks \citep[\eg][]{kang12, pinzke13}. The re-acceleration of a pre-existing population of relativistic electrons injected by nearby radio galaxies is usually invoked to alleviate the high acceleration efficiency required for many relics \citep[\eg][]{botteon16a115, eckert16a2744, vanweeren16toothbrush, hoang18a1240}. To date, the clearest example of an AGN--relic connection is provided by Abell 3411-3412 \citep{vanweeren17a3411}, in which a shock was suggested to be responsible for the AGN distorted radio tail and spectral index flattening at the edge of the relic. The peripheral emission in A781 could resemble this case, provided that future observations will confirm the optical counterpart and shock front.

\subsection{Triple merger in A781}

The detection of discontinuities in the thermal ICM requires that the collision is occurring almost exactly in the plane of the sky, as projection effects could hide the sharp surface brightness and temperature jumps. Therefore, the shock and cold fronts observed in A781 can be used to outline the approximate geometry of the merger. We complemented this information with the temperature and entropy maps shown in Fig.~\ref{fig:icm_maps} (the error maps are reported in Appendix~\ref{app:erro}), which are useful diagnostic tools to search for substructures in the ICM. Maps were produced by fitting a thermal model to the count rates measured in five energy bands from \xmm\ \epic\ Voronoi tessellated images \citep{cappellari03} and requiring a threshold of 400 counts per bin in the $0.5-2.0$ \kev\ band \citep[for more details, see][]{jauzac16}. Reported quantities are projected along the line of sight.

\begin{figure}
 \centering
 \includegraphics[width=.9\hsize]{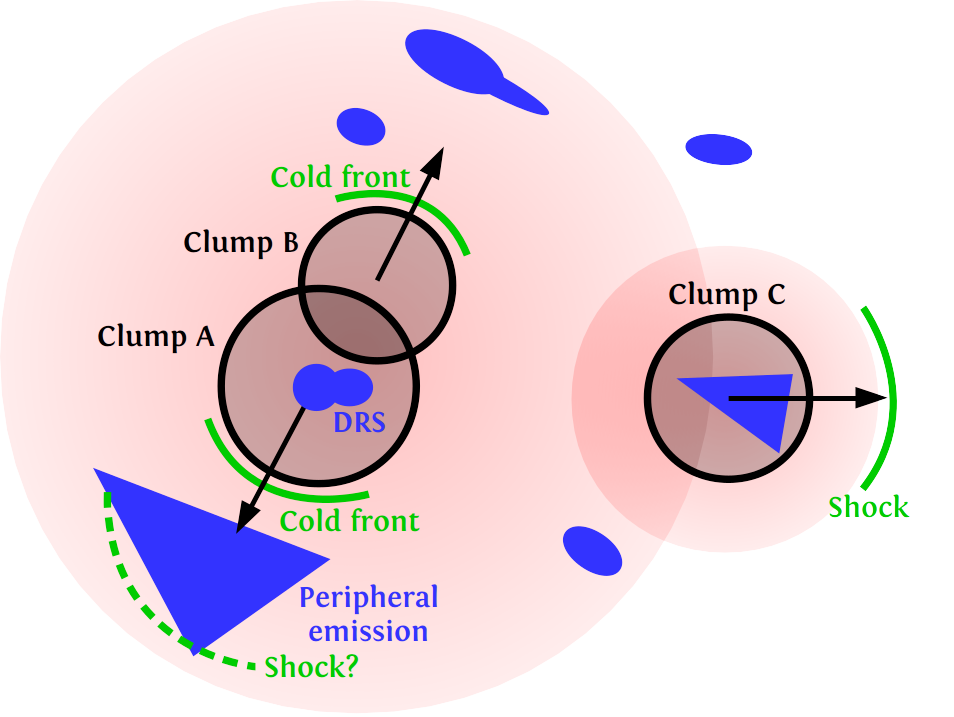}
 \caption{Dynamics of the merger in A781 as suggested from the X-ray data; the diffuse radio sources are sketched in blue while the thermal ICM emission is reported in red (\cf\ Fig.~\ref{fig:rgb}).}
 \label{fig:cartoon}
\end{figure}

From the analysis of the \xmm\ observations, we suggest that A781 is undergoing a triple merger, as sketched in Fig.~\ref{fig:cartoon}. Merger cold fronts usually trace the direction of motion of a cluster core \citep[\eg][]{markevitch02bullet}; hence, the two diametrically opposite cold fronts detected in A781 suggest a collision axis along the NW-SE direction. The presence of two substructures (clump A and clump B) is supported by the low values of entropy and the X-ray contours in Fig.~\ref{fig:icm_maps} (right panel). The two X-ray clumps seem detached (\eg\ Fig.~\ref{fig:rgb}); clump B  likely traces a smaller substructure moving apart from the dominant clump A. The spatial coincidence between the peripheral radio emission high surface brightness spine and bins with $kT \sim 9$ \kev\ in Fig.~\ref{fig:icm_maps} (left panel) could indicate a region heated by the passage of the shock invoked in the previous section to explain the properties of the source observed in the radio band. In addition, the presence of a third sub-cluster (clump C) is highlighted by the X-ray clump of emission to the W and, again, by the low entropy gas in this region (Fig.~\ref{fig:icm_maps}, right panel). This sub-cluster is clearly disturbed as it does not show evidence of an X-ray peak (Fig.~\ref{fig:sectors}). In this respect, we suggest that it is moving towards the W direction and it has already crossed the ICM of clump A+B, rather than infalling into the system. The detection of the shock in the W supports this scenario. This provides an additional merger axis in the E-W direction. Overall, the irregular distribution of temperature with the existence of blobs of hot gas (Fig.~\ref{fig:icm_maps}, left panel) is in agreement with a complex merger dynamics as that described above. \\
\indent
The tentative dynamics of the merger outlined above is based on the features observed in X-ray wavelengths. Recently, \citet{golovich18arx} have presented an optical analysis of A781 that supports the triple merger scenario. As pointed out by these authors, it is worth noting that on larger scales the merger could be even more complex because of the existence of the ``Middle'' cluster, located at a similar redshift of A781 (Fig.~\ref{fig:smoothed}).

\section{Conclusions}

We presented a joint radio/X-ray analysis of the cluster chain Abell 781 using new \lofar\ data and reanalysing archival \gmrt\ and \xmm\ observations. We focussed on the main merging component of the complex, for which the presence of non-thermal emission in the ICM was already investigated in the literature. Our results can be summarised as follows. \\
\indent
Firstly, the nature of the peripheral radio emission in the SE of A781 remains uncertain. We suggested that this source results from the interaction between a weak shock and a radio galaxy. This scenario could explain its unusual morphology and spectral index steepening towards the cluster centre. Future optical follow-up and numerical simulations are required to clarify the origin of the source. \\
\indent
Secondly, we proposed a tentative interpretation of the dynamics of the merger occurring in A781 where three substructures are involved. We detected two cold fronts and a shock front; these were used to delineate the motion of the three mass clumps. The two diametrically opposite cold fronts indicate a merger axis in the SE-NW direction, while the presence of a third substructure moving towards the W and preceding a shock suggests another merger axis in the E-W direction. Three entropy clumps are also observed in the entropy map of A781. \\
\indent
Lastly, our results from the new \lofar\ data and the reanalysis of the archival \gmrt\ observations do not indicate evidence of the radio halo in A781 \citep[in agreement with][]{venturi08, venturi11} and in the other clusters of the chain. We placed an upper limit on the diffuse radio emission a factor of 2 below the $P_{1.4\,\rm{GHz}}-\mfive$ relation of \citet{cassano13}. This limit is not extremely deep due to the presence of artefacts around the bright radio galaxy at the centre of A781.

\begin{acknowledgements}
We thank G.~Bernardi, V.~Cuciti, and K.~Duncan for useful discussions. ABon acknowledges support from the ERC-Stg 714245 DRANOEL. RJvW acknowledges support from the VIDI research programme with project number 639.042.729, which is financed by the Netherlands Organisation for Scientific Research (NWO). FdG is supported by the VENI research programme with project number 639.041.542, which is financed by the Netherlands Organisation for Scientific Research (NWO). The LOFAR group in Leiden is supported by the ERC Advanced Investigator programme New-Clusters 321271. AD and CD acknowledge support by the BMBF Verbundforschung under the grant 05A17STA. This paper is based (in part) on data obtained with the International LOFAR Telescope (ILT) under project code LC6\_015. LOFAR \citep{vanhaarlem13} is the LOw Frequency ARray designed and constructed by ASTRON. It has observing, data processing, and data storage facilities in several countries, which are owned by various parties (each with their own funding sources), and are collectively operated by the ILT foundation under a joint scientific policy. The ILT resources have benefitted from the following recent major funding sources: CNRS-INSU, Observatoire de Paris and Universit\'{e} d'Orl\'{e}ans, France; BMBF, MIWF-NRW, MPG, Germany; Science Foundation Ireland (SFI), Department of Business, Enterprise and Innovation (DBEI), Ireland; NWO, The Netherlands; The Science and Technology Facilities Council, UK; Ministry of Science and Higher Education, Poland. Part of this work was carried out on the Dutch national e-infrastructure with the support of the SURF Cooperative through grant e-infra 160022 \& 160152.  The LOFAR software and dedicated reduction packages on \url{https://github.com/apmechev/GRID\_LRT} were deployed on the e-infrastructure by the LOFAR e-infragroup, consisting of J.~B.~R.~Oonk (ASTRON \& Leiden Observatory), A.~P.~Mechev (Leiden Observatory), and T.~Shimwell (ASTRON) with support from N.~Danezi (SURFsara) and C.~Schrijvers (SURFsara). We thank the staff of the GMRT who made these observations possible. The GMRT is run by the National Centre for Radio Astrophysics of the Tata Institute of Fundamental Research. This work is also based on observations obtained with \xmm, an ESA science mission with instruments and contributions directly funded by ESA Member States and NASA. This research made use of APLpy, an open-source plotting package for Python hosted at \url{http://aplpy.github.com}.
\end{acknowledgements}

%
%

\bibliographystyle{aa}
\bibliography{library.bib}

\begin{appendix}

\section{Error maps}\label{app:erro}

\begin{figure}[h!]
 \centering
 \includegraphics[width=\hsize]{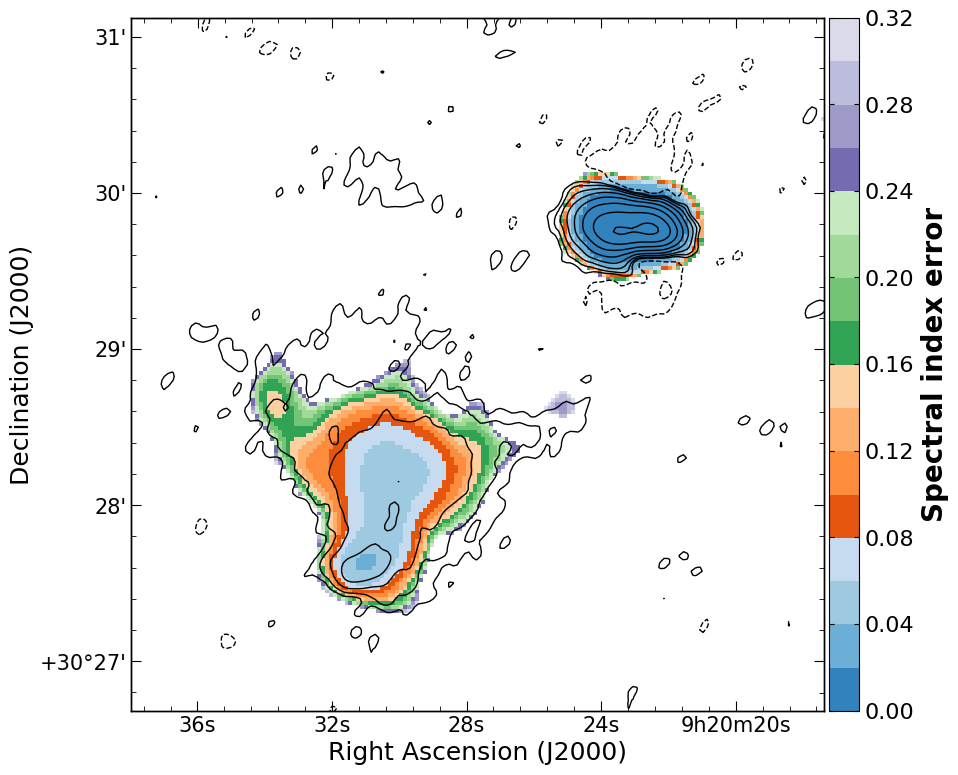}
 \caption{Spectral index error map corresponding to Fig.~\ref{fig:alpha_map}.}
 \label{fig:alpha_error}
\end{figure}

\begin{figure*}[h!]
 \centering
 \includegraphics[width=.45\textwidth,valign=c]{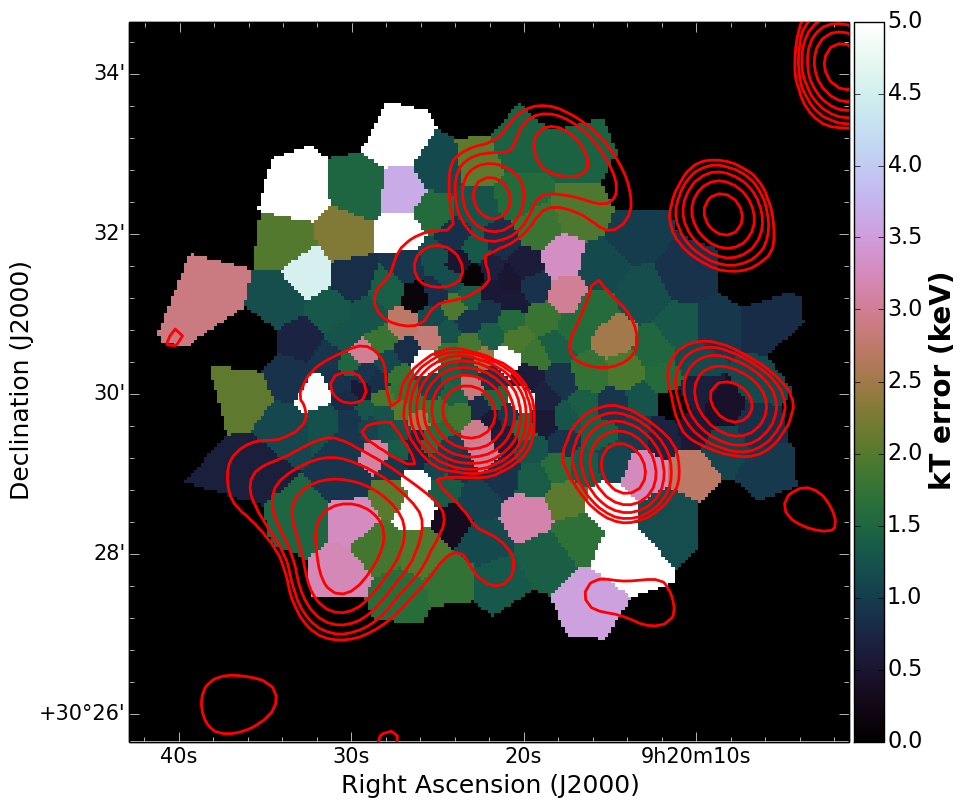}
 \includegraphics[width=.45\textwidth,valign=c]{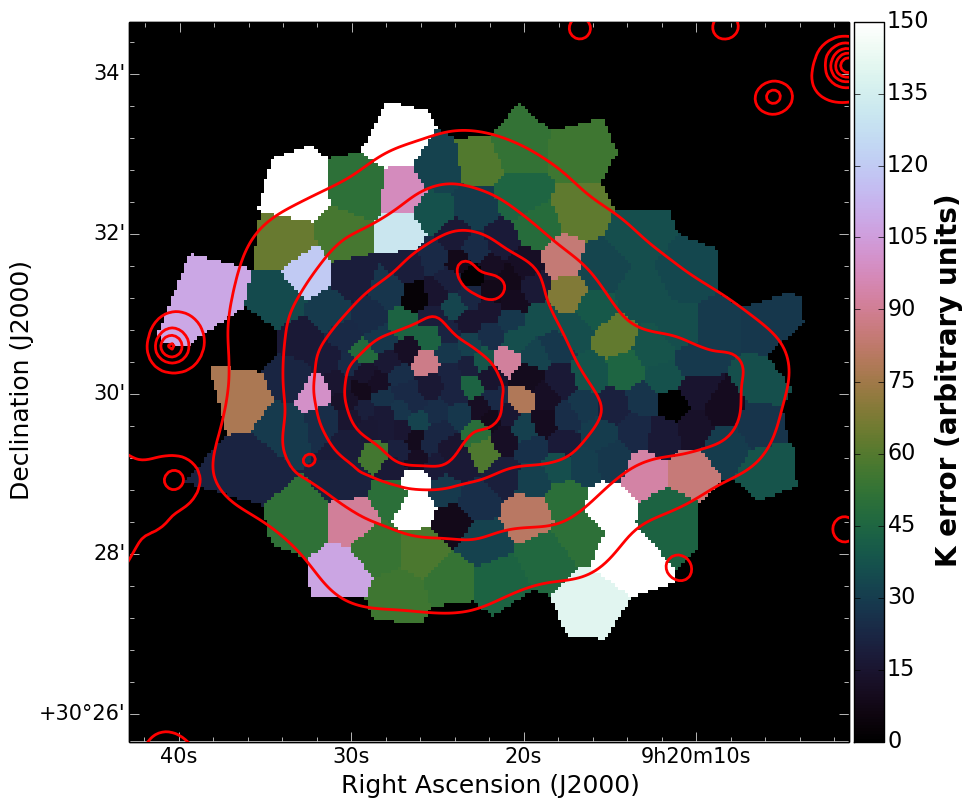}
 \caption{Temperature (\textit{left}) and entropy (\textit{right}) error maps corresponding to Fig.~\ref{fig:icm_maps}.}
 \label{fig:thermo_error}
\end{figure*}

Error maps for the spectral index (Fig.~\ref{fig:alpha_error}), temperature, and entropy (Fig.~\ref{fig:thermo_error}).

\end{appendix}

\end{document}